\documentclass[12pt]{article}
\usepackage{amssymb}
\usepackage{graphicx}
\usepackage{amsmath,amscd}
\def\be{\begin{eqnarray}}
\def\ben{\begin{eqnarray*}}
\def\ee{\end{eqnarray}}
\def\een{\end{eqnarray*}}
\def\Tr{{\rm Tr}}

\def\D{\mathcal{D}}

\def\=:{=\hspace{-.7em}\raisebox{1.1ex}{.}\hspace{.1em}\raisebox{-0.2ex}{.} }
\newcommand{\NF}{N_{\rm F}}
\newcommand{\NC}{N_{\rm C}}

\def\N{{\cal N}}

\newcommand {\beq}{\begin{eqnarray}}
\newcommand {\eeq}{\end{eqnarray}}

\makeatletter
\@addtoreset{equation}{section}

\renewcommand{\thefootnote}{\fnsymbol{footnote}}
\makeatother

\makeatletter
\newcommand{\thetablename}{Table}
\def\fnum@table{\thetablename\ \thetable}
\makeatother

\begin{document}
\thispagestyle{empty}
\begin{flushright}
TIT/HEP--538 \\
{\tt hep-th/0506135} \\
June, 2005 \\
\end{flushright}
\vspace{3mm}

\begin{center}
{\Large \bf 
Webs of Walls
}
\\[10mm]
\vspace{5mm}

\normalsize
{\large 
Minoru~Eto}\!\!
\footnote{\it  e-mail address: 
meto@th.phys.titech.ac.jp
}, 
  {\large 
Youichi~Isozumi}\!\!
\footnote{\it  e-mail address: 
isozumi@th.phys.titech.ac.jp
}, 
  {\large 
Muneto~Nitta}\!\!
\footnote{\it  e-mail address: 
nitta@th.phys.titech.ac.jp
}, \\
  {\large 
 Keisuke~Ohashi}\!\!
\footnote{\it  e-mail address: 
keisuke@th.phys.titech.ac.jp
} 
~and~~  {\large 
Norisuke~Sakai}\!\!
\footnote{\it  e-mail address: 
nsakai@th.phys.titech.ac.jp
} 

\vskip 1.5em

{ \it Department of Physics, Tokyo Institute of 
Technology \\
Tokyo 152-8551, JAPAN  
 }
\vspace{12mm}

\abstract{
Webs of domain walls are constructed as 1/4 BPS states 
in $d=4$, ${\cal N}=2$ supersymmetric $U(N_{\rm C})$ 
gauge theories with $N_{\rm F}$ hypermultiplets 
in the fundamental representation. 
Web of walls can contain any numbers of external legs and 
loops like $(p,q)$ string/5-brane webs. 
We find the moduli space ${\cal M}$ 
of a $1/4$ BPS equation for wall webs
to be the complex Grassmann manifold. 
When  moduli spaces of $1/2$ BPS states (parallel walls) 
and the vacua are removed from ${\cal M}$, the non-compact 
moduli space of genuine $1/4$ BPS wall webs is obtained. 
All the 
solutions are obtained explicitly and exactly
in the strong gauge coupling limit.
In the case of Abelian gauge theory, we work out 
the correspondence between configurations of wall web 
and the moduli space ${\bf C}P^{\NF-1}$. 
}
\end{center}
\vfill
\newpage
\setcounter{page}{1}
\setcounter{footnote}{0}
\renewcommand{\thefootnote}{\arabic{footnote}}



\section{Introduction}

Dirichlet-branes (D-branes) have been essential tools to study 
non-perturbative aspects of string theories and M-theory 
since their discovery~\cite{Polchinski:1998rq}. 
Domain walls may give a field theoretical realization 
of D-branes. 
For instance, like D-branes, domain walls 
become Bogomol'nyi-Prasad-Sommerfield (BPS) states 
preserving the half of supersymmetry (SUSY) 
in SUSY field theories, 
and their tensions saturate the Bogomol'nyi bound~\cite{Witten:1978mh,Cvetic:1991vp}.
They appear as $(1/2,1/2)$ tensorial central charges 
in the SUSY algebras~\cite{Dvali:1996bg}. 
BPS domain walls were extensively 
studied in ${\cal N}=1$ Wess-Zumino models and 
SUSY gauge theories with four 
supercharges~\cite{Chibisov:1997rc}--\cite{N=1wall-moduli},
and ${\cal N}=2$ SUSY gauge theories (with eight supercharges) 
and their associated hyper-K\"ahler 
nonlinear sigma models~\cite{N=2walls}--\cite{SakaiYang}. 
Coupling these walls to ${\cal N}=1$ SUGRA was discussed in 
\cite{Cvetic:1991vp,SUGRA1} and ${\cal N}=2$ SUGRA was in
\cite{SUGRA}. 
Besides walls, vortices in non-Abelian gauge theory 
have also been extensively studied recently \cite{vortices}.
It has been found that in 
$d=4$, ${\cal N}=2$ SUSY 
$U(\NC)$ gauge theories with matter hypermultiplets
(or their associated hyper-K\"ahler nonlinear sigma models), 
vortices (or lumps) as strings can end on 
a wall \cite{Gauntlett:2000de} 
and can be stretched between parallel walls \cite{INOS3},  
like configurations made of strings and D-branes.  
String-wall junction has been studied further 
in \cite{Sakai:2005sp,Auzzi:2005yw}.
These models admit more varieties of composite solitons 
like a monopole (0-brane) attached 
by vortices (strings)~\cite{vm,Eto:2004rz}\footnote{
A similar configuration was discussed in \cite{Kneipp}.}
and a string intersection~\cite{Naganuma:2001pu}. 
Lifting up to $d=5$, 
instanton attached by vortices and intersecting vortices 
carrying instanton charge were found~\cite{Eto:2004rz}.
These composite solitons in field theory may be regarded as 
toy models of composite configurations of several 
D-branes in string theory~\cite{brane-config}.
\begin{figure}[ht]
\vspace*{-2cm}
\begin{center}
\includegraphics[width=15cm]{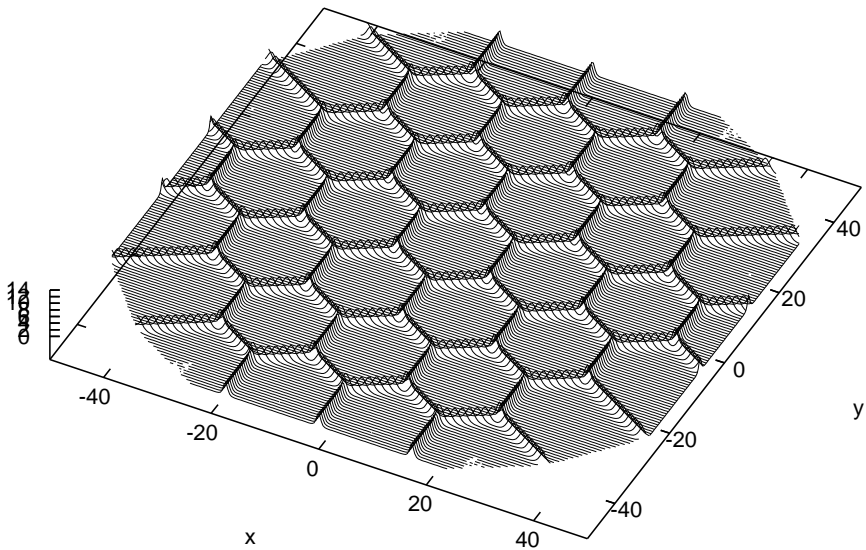}
\vspace*{-1cm}
\caption{\small{Honeycomb web of domain walls. This web divides
37 vacua and has 18 external legs and 19 internal faces.
The moduli space corresponds to ${\bf C}P^{36}$ whose dimension is 72.}}
\label{hachinosu}
\end{center}
\end{figure}

We can pursue similarities further by moving
to junctions of branes.
Type IIB string theory is 
invariant under S-duality  
which exchanges a fundamental string and a D-string 
or a NS5-brane and a D5-brane. 
Several $(p,q)$ strings with NS-NS charge $p$ and 
R-R charge $q$ 
can make a junction balancing tensions 
at a junction point~\cite{Aharony:1996xr}. 
The $(p,q)$ string webs stretched between multiple 
parallel D$3$-branes 
are regarded as 1/4 BPS dyon in the D3-brane 
effective gauge theory~\cite{1/4dyon,Bergman:1998gs}. 
Several $(p,q)$ 5-branes can also form 
a junction~\cite{5-brane}.
Type IIB string theory admits more general configurations 
made of several connected strings or 5-branes, 
called {\it string webs} or {\it 5-brane webs}, respectively.
They are 1/4 BPS states and give planer diagrams.  
On the other hand domain walls in field theory also can form  
a junction~\cite{Abraham:1990nz}.  
It was shown \cite{Gibbons:1999np,Carroll:1999wr} that 
wall junctions preserve a quarter
of SUSY and therefore 
are 1/4 BPS states 
in $d=4$, ${\cal N}=1$ SUSY field theories. 
The energy density is bounded 
from below by two types of 
central charge densities ${\cal Z}_\alpha$ ($\alpha=1,2$) and ${\cal Y}$ 
with ${\cal Z}_\alpha$ for walls perpendicular to 
the $\alpha$-th direction of two co-dimensions 
and ${\cal Y}$ for a junction\footnote{
The central charge density
${\cal Y}$ has been discussed previously in 
the context of vortices \cite{Chibisov:1997rc}.
}. 
Subsequently a number of works 
appeared 
on domain wall 
junctions~\cite{Oda:1999az}--\cite{Kakimoto:2003zu}.
An exact solution of a ${\bf Z}_3$ symmetric 
wall junction was constructed in the Wess-Zumino 
model and its junction charge ${\cal Y}$ was found to 
be negative \cite{Oda:1999az}, which was recognized 
to be natural later \cite{Shifman:1999ri}.
Exact solutions for ${\bf Z}_n$ symmetric wall junctions 
were also found in nonlinear sigma models \cite{Naganuma:2001br}.  
As exact solutions in ${\cal N}=2$ SUSY theories, 
a domain wall intersection was obtained 
in hyper-K\"ahler nonlinear sigma models \cite{Gauntlett:2000bd}, 
and a ${\bf Z}_3$ symmetric wall junction was presented in 
$d=4$, ${\cal N}=2$ SUSY $U(1)$ gauge theories with three 
hypermultiplets in the fundamental representation 
\cite{Kakimoto:2003zu}.
Embedding a wall junction to SUGRA was also discussed 
\cite{Carroll:1999mu}. 
In either case, only single junctions have been obtained 
as exact solutions so far. 
Domain walls are, however, expected to make a 
network when many junctions are connected together 
\cite{network} like webs, although no explicit solutions 
are available yet.
Such domain wall networks are also studied in 
cosmology~\cite{cosmology,VS} typically using 
numerical simulations instead of explicit solutions.
Although a qualitative discussion has been made 
on partial moduli for a simple junction  
in the literature \cite{Carroll:1999wr}, 
no quantitative treatment 
of a complete moduli space was available so far. 

The purpose of this paper is to 
construct all 
the solutions of domain wall webs as 1/4 BPS states 
in $d=4$, ${\cal N}=2$ SUSY $U(\NC)$ 
gauge theories with $\NF (> \NC)$ hypermultiplets 
in the fundamental representation with complex masses. 
The energy density is saturated by central charge densities 
${\cal Z}_\alpha$ ($\alpha=1,2$) and ${\cal Y}$ 
as usual~\cite{Gibbons:1999np,Carroll:1999wr}. 
We find that 
the moduli space of the $1/4$ BPS equation for wall web 
in this theory 
to be the complex Grassmann manifold 
$G_{\NF,\NC} \simeq SU(\NF)/[SU(\NF-\NC) \times SU(\NC) \times U(1)]$. 
Moreover exact solutions are obtained 
in the strong gauge coupling limit. 
Our wall webs contain several external legs and loops 
whose maximal numbers are determined by $\NF$ and $\NC$.
An amusing illustrative example of the 
exact solution is displayed in Fig.~\ref{hachinosu}. 
In the case of the Abelian gauge theory, we work out
the correspondence between configurations of wall webs and
the moduli space ${\bf C}P^{\NF-1}$.
In that case we find that the web configurations are 
naturally expressed by the grid diagrams 
in the complex $\Sigma$ plane ($\Sigma$ is the
complex scalar field in the vector multiplet), 
which are dual to the web diagrams in the configuration space. 
While vacua and 1/2 BPS walls correspond to the vertices and
segments of the grid diagram, junctions correspond to the
faces.  
We find that 
the wall charges 
are proportional to the lengths of edges of the grid diagram 
whereas the junction charge is to the area of the grid diagram. 
The grid diagram has been found in the context of 
$(p,q)$ string/5-brane webs in the superstring theory.
Our wall webs have a strong similarity with 
string/5-brane webs when we identify $(p,q)$ charge 
with the wall central charges $(Z_1,Z_2)$.  
Therefore 
these solutions can be called $(p,q)$ {\it wall webs}.

The moduli space of 1/4 BPS solutions exhibits 
an interesting structure. 
To see it let us first recall our previous result on 
the moduli space of 1/2 BPS domain walls~\cite{INOS1,INOS2}. 
In this case, all topological sectors with various dimensions 
are patched together to form the total moduli space, 
which is again the complex Grassmann manifold. 
There zero wall sectors with zero dimension 
(isolated points) corresponding to vacua are added to make 
the total moduli space compact. 
In the case of the moduli space for solutions of 
1/4 BPS equations,  
the moduli space ${\cal M}_{1/4}$ for genuine 1/4 BPS states 
is obtained by removing the moduli space ${\cal M}_{1/2}$ for 
1/2 BPS parallel walls and 
the moduli space ${\cal M}_{1/1}$ for vacua 
from the total moduli space ${\cal M}_{\rm tot.} \simeq G_{\NF,\NC}$ 
\begin{equation}
{\cal M}_{\rm tot.} ={\cal M}_{1/4} + {\cal M}_{1/2} + 
{\cal M}_{1/1}. 
\end{equation}
Moreover the moduli space ${\cal M}_{1/4}$ for 
genuine 1/4 BPS states is further decomposed into 
topological sectors according to the number of junction 
points.

This paper is organized as follows. 
In Sec.~\ref{sect2}, 1/4 BPS equations for junctions are 
obtained and solved. 
The total moduli space for the wall webs are given 
and the exact solutions in the strong coupling limit 
is presented. 
In Sec.~\ref{sect3} we study the explicit relationship between 
the moduli parameters 
and the configuration of the wall webs 
in the case of Abelian gauge theory with $\NC=1$. 
The junction charge ${\cal Y}$ is shown to be always 
negative in the Abelian gauge theory. 
In Sec.~\ref{sect4}, we discuss 
wall junction in more general gauge theories. 
We present a method to estimate the shape 
of 1/4 BPS wall junctions once 1/2 BPS parallel wall 
configurations are known in a corresponding 
theory with real masses. 
Sec.~\ref{sect5} is devoted to a discussion. 
In Appendix A, we show that 
no additional moduli parameters appear from gauge fields  
in the case of  the $U(1)$ gauge theory.

\section{1/4 BPS equations and their solutions \label{sect2}}
We start with 3+1 dimensional\footnote{
Wall junctions require complex mass parameters for 
hypermultiplets, which are available only in dimensions 
$d \le 3+1$ as noted in Ref.~\cite{Kakimoto:2003zu}. 
}
 $\N = 2$ supersymmetric 
$U(\NC)$ gauge theory with $\NF(>\NC)$ massive hypermultiplets
in the fundamental representation. 
The physical fields contained in this model are a gauge 
field $W_\mu\ (\mu=0,1,2,3)$, real adjoint scalars 
$\Sigma_\alpha\ (\alpha=1,2)$ and gaugino $\lambda^i$
in the vector multiplet, and $\NF$ complex 
doublets of scalars $H^{iAr}$ 
$(r=1,2,\cdots,\NC,\ A=1,2,\cdots,\NF,\ i=1,2)$ 
and its superpartners $\psi^{Ar}$ in the hypermultiplets.
We express $\NC\times\NF$ matrix of the hypermultiplets by 
$H^i$. 
With the metric $\eta_{\mu\nu}=(+1,-1,-1,-1)$, we obtain 
the bosonic Lagrangian 
\be
{\cal L} 
&=&
\Tr\left[
-\frac{1}{2g^2}F_{\mu\nu}F^{\mu\nu}
+ \frac{1}{g^2}\sum_{\alpha=1}^2
\D_\mu\Sigma_\alpha\D^\mu\Sigma_\alpha 
+ \D_\mu H^i\left(\D^\mu H^i\right)^\dagger
\right] - V,
\label{lag}\\
V &=& \Tr\left[\frac{1}{g^2}\sum_{a=1}^{3}\left(Y^a\right)^2
+ \sum_{\alpha=1}^2\left(H^i M_\alpha - \Sigma_\alpha H^i\right)
\left(H^i M_\alpha - \Sigma_\alpha H^i\right)^\dagger
- \frac{1}{g^2}\left[\Sigma_1,\Sigma_2\right]^2
\right],
\label{pot}
\ee
with diagonal mass matrices 
$M_1={\rm diag}\left(m_1,m_2,\cdots,m_{\NF}\right)$ and 
$M_2 = {\rm diag}\left(n_1,n_2,\cdots,n_{\NF}\right)$. 
Here we define 
$
Y^a \equiv \frac{g^2}{2}
\left(c^a{\bf 1}_{\NC} - (\sigma^a)^j{_i}H^i(H^j)^\dagger\right)
$
with $c^a$ the Fayet-Iliopoulos (FI) parameters.
In the rest of this paper we choose the FI parameters as $c^a=(0,0,c>0)$ 
by using $SU(2)_R$ rotation without loss of generality. 
The covariant derivatives and the field strength are defined by
 $\D_\mu \Sigma_\alpha = \partial_\mu \Sigma_\alpha 
+ i[W_\mu,\Sigma_\alpha]$, 
 $\D_\mu H^i = \partial_\mu H^i + iW_\mu H^i$ and 
 $F_{\mu\nu}=-i[\D_\mu,\D_\nu]=
 \partial_\mu W_\nu -\partial_\nu W_\mu +i[W_\mu, W_\nu]$,  
respectively. 
The supertransformation for spinors is 
\be
\delta \lambda^i &=& \left(
\frac{1}{2}\gamma^{\mu\nu} F_{\mu\nu}
+ \gamma^\mu\D_\mu \Sigma_1 + i\gamma^5\gamma^\mu\D_\mu\Sigma_2
- \gamma^5\left[\Sigma_1,\Sigma_2\right]\right)\varepsilon^i
+ iY^a(\sigma^a)^i{_j}\varepsilon^j,\\
\delta \psi &=& \sqrt 2
\left[-i\gamma^\mu\D_\mu H^i
+ \left(\Sigma_1 H^i - H^iM_1\right)
-i\gamma^5\left(\Sigma_2 H^i - H^i M_2\right)
\right]\epsilon_{ij}\varepsilon^j,
\ee
with $\epsilon_{12} = \epsilon^{12} = 1$.

When we turn off all the mass parameters, 
the vacuum manifold of the above model
becomes complex Grassmann manifold
$T^* G_{\NF,\NC}$ with its size $c$.
Once mass parameters $m_A+in_A$ in $M\equiv M_1+iM_2$ are turned on 
and are chosen to be nondegenerate, 
almost all points of the vacuum manifold are lifted and
only $_{\NF}\!C_{\!\NC} = \NF !/(\NC !(\NF-\NC)!)$ points 
on the base manifold $G_{\NF,\NC}$ 
remain as the discrete SUSY vacua~\cite{ANS}. 
Color and flavor are locked in these vacua. 
Each vacuum is characterized 
by a set of $\NC$ different flavor indices 
$\{A_{1}, \cdots,A_{N_{\rm C}}\}$ 
out of $N_{\rm F}$ flavors 
corresponding to the non-vanishing hypermultiplets as 
$H^{1rA} = \sqrt c \delta^{A_r}{_A},\ H^{2rA} = 0$. 
This vacuum is labeled by $\langle A_1A_2\cdots A_{\NC}\rangle $. 
Here we suppress phase factors by using global gauge 
transformations.
In these vacua, 
the complex adjoint scalar $\Sigma \equiv \Sigma_1+i\Sigma_2$ 
have the vacuum expectation value determined by 
the mass parameters of the corresponding flavors 
\begin{eqnarray}
\langle \Sigma \rangle_{\langle A_1\cdots A_{\NC}\rangle} 
= {\rm diag}\left(m_{A_1}+in_{A_1},m_{A_2}+in_{A_2},\cdots,m_{A_{\NC}}+in_{A_{\NC}}\right).
\end{eqnarray}

Let us next derive 1/4 BPS equations for string webs by usual 
Bogomol'nyi completion of the energy density. 
We ignore $H^2$ below because 
it always vanishes for the following
1/4 BPS equations. 
In the following we simply denote $H^1 \equiv H$. 
We consider static configurations which are independent on $x^3$ 
$(\partial_0=\partial_3=0)$
and set $W_0 = W_3 = 0$.
Then the Bogomol'nyi completion is performed as\footnote{
Here and in the following we denote spacial indices of 
codimensions of the wall web by $\alpha=1, 2$ using the same 
notation as the indices for the adjoint scalar 
$\Sigma_\alpha$. 
Although $\Sigma_\alpha$ may be understood as 
two extra-dimension components of gauge fields 
of compactified six-dimensional SUSY gauge theory, no 
confusion hopefully arise with this notation. 
}
\be
{\cal E} 
&=& \Tr\bigg[
\frac{1}{g^2}\left(F_{12} - i \left[\Sigma_1,\Sigma_2\right]\right)^2
+ \frac{1}{g^2}\left(\D_1\Sigma_2 - \D_2\Sigma_1\right)^2
+ \frac{1}{g^2}\left(\D_1\Sigma_1 + \D_2\Sigma_2 - Y^3\right)^2\nonumber\\
 &+& \sum_{\alpha=1,2}\left(\D_\alpha H - HM_\alpha + \Sigma_\alpha H\right)
\left(\D_\alpha H - HM_\alpha + \Sigma_\alpha H\right)^\dagger
\bigg] + \sum_{\alpha=1,2}\partial_\alpha J_\alpha
+ {\cal Y} + {\cal Z}_1 + {\cal Z}_2\nonumber\\
&\ge& {\cal Y}+ {\cal Z}_1 + {\cal Z}_2 + \sum_{\alpha=1,2}\partial_\alpha J_\alpha,
\label{energy_comp}
\ee
where the central charge densities are of the form
\be
{\cal Y} \equiv \frac{2}{g^2}
\partial_\alpha\Tr\left(\epsilon^{\alpha\beta}
\Sigma_2\D_\beta\Sigma_1\right),\quad
{\cal Z}_1 \equiv c \partial_1 \Tr \Sigma_1,\quad
{\cal Z}_2 \equiv c \partial_2 \Tr \Sigma_2
\label{cetral charge}
\ee
Here we define
$J_\alpha \equiv {\rm Tr }\Big[ 
H(M_\alpha H^\dagger - H^\dagger \Sigma_\alpha)\Big] $ 
in the last line of Eq.~(\ref{energy_comp}).
Notice that this can't have any contribution to
topological charges after integrating over the $x^1$-$x^2$ plane.
The charge density ${\cal Z}$ counts the domain wall charge 
$\pi_0$ and the charge density ${\cal Y}$ counts 
the junction charge $\pi_1$ in the $x^1$-$x^2$ plane.

From the condition that the above energy bound is saturated, 
the BPS equations for domain wall webs can be obtained as 
\be
&&F_{12} = i \left[\Sigma_1,\Sigma_2\right],\quad
\D_1\Sigma_2 = \D_2\Sigma_1,\quad
\D_1\Sigma_1 + \D_2\Sigma_2 = Y^3,
\label{bps_eq1}\\
&&\D_1 H = HM_1 - \Sigma_1 H,\quad
\D_2 H = HM_2 - \Sigma_2 H.
\label{bps_eq2}
\ee
These BPS equations can be also derived from the 
conservation condition for the part of SUSY defined by 
the projections 
$\Gamma_*\varepsilon = \varepsilon$ with the following 
gamma matrices for $\Gamma_*$ 
\be
\Gamma_{\rm w} = \gamma^1\otimes i\sigma^3,\quad
\Gamma_{\rm w'} = -i\gamma^2\gamma^5\otimes i\sigma^3,\quad
\Gamma_{\rm j} = \gamma^0\gamma^3\otimes{\bf 1}_2.
\label{eq:gamma-proj}
\ee
These three gamma matrices commute 
with each other and the product of any two gives the 
remaining one. 
Therefore we conclude that
the above BPS Eqs. (\ref{bps_eq1}) and (\ref{bps_eq2}) 
preserve 1/4 SUSY.
Note that because of a property 
${\gamma ^{12}}\varepsilon^i = -i\gamma ^5\varepsilon^i $
on the projected Killing spinor, 
we can rotate the configuration 
in $z\equiv x^1+ix^2$ 
plane with accompanying chiral rotation\footnote{
The projection (\ref{eq:gamma-proj}) 
defining the conserved SUSY relates the 
$\alpha=1, 2$ indices in real space $x^\alpha$ 
and $\Sigma_\alpha$. Of course, the different SUSY are 
conserved when we perform only the spatial rotation 
without the accompanying chiral rotation. 
} 
\begin{eqnarray}
\left(z,\Sigma,M\right) \to e^{i\theta}\left(z,\Sigma,M\right), 
\quad
\left(\lambda^i,\psi\right) \to 
e^{-i\frac{\theta}{2}\gamma^5}
e^{-\frac{\theta}{2}\gamma^{12}}\left(\lambda^i,\psi\right),
\label{eq:rotation}
\end{eqnarray} 
maintaining the same combination of 1/4 SUSY. 
Then the formulae for the 1/4 BPS equations and the central 
charges remain intact.

The BPS Eqs.(\ref{bps_eq1}) and (\ref{bps_eq2}) reduce to 
the 1/2 BPS equations for the non-Abelian parallel walls 
\cite{INOS1,INOS2} when we turn off both the 
$x^2$-dependence and the mass $M_2$. 
Their solutions and the total moduli space were found 
in Ref.~\cite{INOS1}. 
Moreover, the Eqs.(\ref{bps_eq1}) and (\ref{bps_eq2})
can be derived by performing the Scherk-Schwarz (SS) dimensional reduction
once from the other 1/4 BPS system with monopoles, vortices and walls
\cite{INOS3}, and also derived by SS dimensional reduction twice
from another 1/4 BPS system with vortices and instantons \cite{Eto:2004rz}.
The method developed in Refs.~\cite{INOS1,INOS2,INOS3,Eto:2004rz} 
to solve BPS equations can be extended to these 1/4 BPS 
equations (\ref{bps_eq1}) and
(\ref{bps_eq2}).
At the first step we solve the 
first two of Eqs.(\ref{bps_eq1}) and (\ref{bps_eq2}).
Notice that the first two in Eqs.(\ref{bps_eq1}) are an 
integrability condition\footnote{
In fact first two of Eqs.(\ref{bps_eq1}) can be 
rewritten as $\left[\D_1+\Sigma_1,\D_2+\Sigma_2\right]=0$.
}
for 
simultaneous solutions of 
two equations in Eqs.(\ref{bps_eq2}) 
to exist consistently. 
Solutions of Eq.~(\ref{bps_eq2}) are obtained in terms of 
$\NC\times\NC$ 
non-singular matrix $S(x^\alpha)$ as
\be
H = S^{-1}H_0e^{M_1x^1 + M_2x^2},\quad
W_1 - i\Sigma_1 = -iS^{-1}\partial_1 S,\quad
W_2 - i\Sigma_2 = -iS^{-1}\partial_2 S.
\label{bps_sol}
\ee
Here $H_0$ is an $\NC\times\NF$ constant complex 
matrix of rank $\NC$. 
We call $H_0$ the {\it moduli matrix} because it contains 
moduli parameters of solutions as we see below.
Defining a gauge invariant matrix 
\be 
 \Omega\equiv SS^\dagger
\ee
the last of Eqs.(\ref{bps_eq1}) can be written as
\be
{1 \over cg^2}\left[
\partial_1\left(\partial_1\Omega\Omega^{-1}\right)
+ \partial_2\left(\partial_2\Omega\Omega^{-1}\right)
\right]
= {\bf 1}_{\NC} - \Omega_0\Omega^{-1},
\label{master}
\ee
with a ``source'' 
\be 
 \Omega_0 \equiv c^{-1}H_0e^{2(M_1x^1+M_2x^2)}H_0^\dagger.
\ee
We call Eq.~(\ref{master}) the {\it master equation} 
of our 1/4 BPS system (\ref{bps_eq1})
and (\ref{bps_eq2}) since all configurations of
the physical fields $H$, $\Sigma_\alpha$
and $W_\alpha$ are determined (with appropriate 
gauge choice) from the solution $\Omega$.
The master equation (\ref{master}) should be solved
with appropriate boundary conditions which we will discuss 
in the following section.
Notice that the solutions have to approach vacua sufficiently 
far from the walls, then $\Omega$ approaches to $\Omega_0$ there.
The energy density of the BPS wall webs in the right-hand 
side of Eq.~(\ref{energy_comp}) 
can be rewritten in terms of $\Omega$ as
\be
{\cal E}_{\rm BPS} = 
\frac{2}{g^2}\Tr\left[\epsilon^{\alpha\beta}
\partial_\alpha\left(\partial_2\Omega\Omega^{-1}\right)
\partial_\beta\left(\partial_1\Omega\Omega^{-1}\right)
\right] + 
\left(\frac{c}{2} - \frac{1}{2g^2}\partial_\alpha^2\right)
\partial_\alpha^2\log\det\Omega,
\label{energy}
\ee
where we use the relation
\be
\Sigma_\alpha = \frac{1}{2} S^{-1} \partial_\alpha 
\Omega  S^{\dagger -1}.
\label{sigma_omega}
\ee

One of the advantages 
of solving the BPS equation partially using 
the matrix $H_0$ in Eq.~(\ref{bps_sol}) is to identify the moduli 
of the domain wall webs. 
The matrix $H_0$ contains parameters of solutions, 
namely moduli parameters. 
However, $(H_0,S)$ and $(H_0',S')$ related by the 
{\it world-volume symmetry} 
\be
 H_0 \to H_0' = VH_0,\quad S \to S'=VS,
\label{eq:world-volume-sym}
\ee
with $V\in GL(\NC,{\bf C})$ give 
the same configurations for the physical fields. 
Therefore the independent moduli parameters 
are given by the equivalence class defined by 
$(H_0,S) \sim (H_0',S')$. 
Then the {\it total moduli space} which 
is just a topological space that consists of all the parameters in the 
moduli matrix $H_0$ is
the complex Grassmann manifold\footnote{
The total moduli space contains moduli parameters
which change boundary conditions of the solution when
the parameters are changed.
Such modes would be non-normalizable and unphysical if we
consider the effective action of the solitons. 
We will discuss the effective theory of the web in Sec.~\ref{sect5}.}
\be
{\cal M}^{\rm webs}_{\rm tot} \simeq  G_{\NF,\NC} =
\{H_0\ |\ H_0\sim VH_0,\ V\in GL(\NC,{\bf C})\}.
\label{tot_mod}
\ee

To show conclusively that the Grassmann manifold contained 
in the moduli matrix $H_0$ is the total 
moduli space, we need to prove the existence and uniqueness 
of solutions of the master equation (\ref{master}). 
This task has been accomplished 
in the case of 1/2 BPS parallel walls in 
$U(1)$ gauge theories \cite{SakaiYang}. 
In the case of the 1/2 BPS parallel walls in $U(N_{\rm C})$ 
gauge theories, no direct proof is available. 
However, we have shown that the number of moduli 
parameters contained in $H_0$ are necessary and sufficient 
as required by the index theorem \cite{Sakai:2005sp,Eto:2005wf}. 
As for the 1/4 BPS wall webs, 
we prove the uniqueness of the 
master equation (\ref{master}) for the case of the $U(1)$ gauge
theories in Appendix. 
Our proof implies that the master equation (\ref{master}) 
in the $U(1)$ gauge theories does not generate additional 
moduli parameters besides those in the moduli matrix $H_0$.
As for the wall webs in the $U(N_{\rm C})$ gauge theories, 
neither direct proof nor the index theorem are not known yet. 
However, in the strong gauge coupling limit the master 
equation (\ref{master}) reduces to just an algebraic 
equation as we will show, so that we can verify that 
all the moduli parameters are contained 
in the moduli matrix for $U(N_{\rm C})$ as well as $U(1)$ 
gauge theories. 
For those cases where a direct proof or index theorem is not available,  
our result (\ref{tot_mod}) 
is at present based on a conjecture that the solution of the 
master equation exists and is unique.

The total moduli space 
(\ref{tot_mod}) of solutions of the 1/4 BPS 
equations (\ref{bps_eq1}) and (\ref{bps_eq2}) 
is a compact manifold. One might feel a little strange because
ordinary moduli spaces of solitons have non-compact directions; at least 
their translational zero modes give non-compact directions.
Our compact total moduli space exhibits an interesting structure as follows.
To see it let us first recall the fact that solution of
our 1/4 BPS equations (\ref{bps_eq1}) and (\ref{bps_eq2}), namely 
the moduli matrix $H_0$,
contains 1/2 BPS and vacuum states besides 1/4 BPS states.
In other words, the compact total moduli space
$G_{\NF,\NC}$ includes
different sectors, namely 
${\cal M}^{\text{webs}}_{1/4}$ for genuine 1/4 BPS states,
the moduli space ${\cal M}^{\text{walls}}_{1/2}$ for 
1/2 BPS walls and ${\cal M}^{\text{vacua}}_{1/1}$ for discrete SUSY vacua 
\be
\label{eq:tot-mod-quart}
{\cal M}^{\rm webs}_{\rm tot}
\simeq G_{\NF,\NC}
= {\cal M}^{\text{webs}}_{1/4}
\ \bigcup\ 
{\cal M}^{\text{walls}}_{1/2}
\ \bigcup\ 
{\cal M}^{\text{vacua}}_{1/1}.
\ee
Notice that the 1/2 BPS wall sector 
${\cal M}^{\text{walls}}_{1/2}$ consists of 
subspaces for the 1/2 BPS walls which preserve different 
sets of 1/2 supercharges.
After decomposition, each sector of $G_{\NF,\NC}$ in 
Eq.~(\ref{eq:tot-mod-quart})
is in fact non-compact except for 
the vacuum sector.
The union of them, however, form the compact manifold 
when all of them
are appropriately patched together.
Thus the moduli space for genuine 1/4 BPS states 
is obtained by removing the moduli spaces for 
1/2 BPS walls and for vacua 
from the total moduli space $G_{\NF,\NC}$.
Moreover the moduli space ${\cal M}^{\text{webs}}_{1/4}$ is 
further decomposed into 
topological sectors according to the number of junction 
points.
In the next section we will illustrate this 
decomposition of the total moduli space 
into 1/4, 1/2, and 1/1 BPS subspaces using several examples 
in the Abelian model.

This kind of the decomposition of the total moduli space 
$G_{\NF,\NC}$ has 
been already observed in Refs.~\cite{INOS1,INOS2} for the 1/2 BPS
parallel domain walls.
Now, we can reproduce the result for 1/2 BPS walls 
as a special case of the 1/4 BPS webs.
Let us first note that
our construction of solving 1/4 BPS equations 
in terms of the moduli matrix $H_0$ is insensitive
to the changes of mass parameters\footnote{
We always assume that the mass parameters are fully 
non-degenerate.}, 
even though changing the mass parameters means changing 
the theory itself.
As was mentioned in the paragraph below Eq.~(\ref{eq:rotation}),
our 1/4 BPS system of the webs reduces to
the 1/2 BPS system of the parallel walls when we turn off
both the $x^2$ dependence and the mass $M_2$.
In that case all the component walls in the webs become parallel
each other, so that the 1/4 BPS sector 
${\cal M}^{\text{webs}}_{1/4}$ and the 1/2 BPS sector 
${\cal M}^{\text{walls}}_{1/2}$ in Eq.~(\ref{eq:tot-mod-quart})
get together and reproduce new 1/2 BPS sector 
${\cal M}^{\text{parallel walls}}_{1/2}$ 
as the moduli space for the 1/2 BPS parallel walls 
\be
{\cal M}^{\text{webs}}_{1/4}
\ \bigcup\ 
{\cal M}^{\text{walls}}_{1/2}
\ \to\ 
{\cal M}^{\text{parallel walls}}_{1/2}.
\ee
Then the 1/2 BPS parallel walls again form the same total 
moduli space $G_{\NF,\NC}$ when these are combined with 
the vacuum sectors as \cite{INOS1}
\be
\label{eq:tot-mod-half}
{\cal M}^{\rm walls}_{\rm tot} 
\simeq G_{\NF,\NC}
= 
{\cal M}^{\text{parallel walls}}_{1/2}
\ \bigcup\ 
{\cal M}^{\text{vacua}}_{1/1}.
\ee

Let us consider the strong gauge coupling limit
$g^2\to\infty$. In this limit the massive 
gauged linear sigma model (\ref{lag}) and (\ref{pot}) 
reduces to the 
massive nonlinear sigma model with $T^* G_{\NF,\NC}$ 
as its target space \cite{LR,ANS}.
The fields in the vector multiplet reduce to 
auxiliary fields which can be expressed in terms of the 
hypermultiplets using their equations of motion:
\be
W_\alpha = -ic^{-1} H\partial_\alpha H^\dagger,\quad
\Sigma_\alpha = c^{-1}HM_\alpha H^\dagger.
\label{aux}
\ee
Since the left hand side of the master equation (\ref{master}) 
vanishes in the strong coupling limit, the right-hand 
side gives just an algebraic equation 
\be
 \Omega^{g \to \infty} = \Omega_0 
= c^{-1}H_0e^{2(M_1x^1+M_2x^2)}H_0^\dagger. 
\ee
$S$ can be calculated from $\Omega$ 
by fixing a gauge,
and the configuration is exactly obtained 
by Eq.~(\ref{bps_sol}). 
In the case of Abelian gauge theory ($\NC=1$) we find 
configurations of scalar fields up to gauge symmetry as 
\be
H^A = \sqrt{c}\,\frac{H_0^Ae^{m_Ax^1+n_Ax^2} }
{\sqrt{\sum_{B=1}^{\NF}
|H_0^B|^2e^{2(m_Bx^1+n_B x^2)}}} .
\label{exact}
\ee  
Note that the central charge ${\cal Y}$ vanishes in this 
limit, so that only ${\cal Z}$ contributes tension of wall 
webs.
We would like to stress that these solutions contain all 
the exact solutions of 1/2 BPS domain walls and 1/4 BPS 
domain wall webs in the massive $T^* G_{\NF,\NC}$ nonlinear 
sigma model.
They can have any number of external walls and internal 
loops in the webs.
In the next section 
we will show several exact solutions of webs 
in Abelian gauge theory 
to give examples.

\section{Abelian domain wall webs \label{sect3}}

In this section we will investigate several fundamental 
properties about
the 1/4 BPS wall junctions and its webs in 
SUSY gauge theories.
For simplicity,  we consider the
Abelian gauge theory $(\NC = 1)$ in 
the following. 
In this case, the moduli matrix $H_0$ 
is an $\NF$ component complex vector. 
As was mentioned in the previous section, $H_0$ should be 
identified with $V H_0$ where $V$ is a complex number. 
Therefore these moduli parameters in $H_0$ are 
the homogeneous coordinate of the complex projective space 
${\bf C}P^{\NF-1}$.
Each point in the moduli space ${\bf C}P^{\NF-1}$ 
expresses a configuration of a wall web, parallel walls 
or vacuum in the real space $x^1$-$x^2$.

We will examine properties of 1/4 BPS wall webs and 
give some examples of the webs in the following subsections
without restricting us to the strong gauge coupling limit. 
Although we cannot solve the master equation 
(\ref{master}) explicitly in the case of finite 
gauge couplings, we can 
clarify several fundamental properties, 
for example tension, position and orientation 
of component walls, 
from informations encoded in the moduli matrix $H_0$.
We will use the infinite gauge coupling limit merely 
to illustrate the wall webs explicitly. 

\subsection{Domain wall}

Before studying 1/4 BPS domain wall webs, 
let us begin with a brief review of the 1/2 BPS single wall 
in the model with relatively real masses for the 
hypermultiplets.
Although this was already studied in detail 
in Refs.~\cite{INOS1,INOS2},
it should be useful to understand 
the structure of domain wall webs 
in the following of this paper.

\begin{figure}[ht]
\begin{center}
\includegraphics[height=4cm]{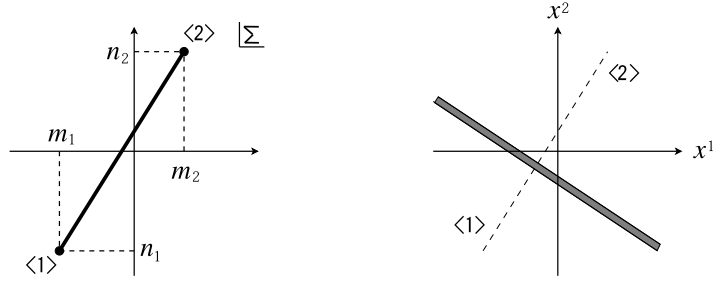}
\caption{\small{1/2 BPS domain wall with a complex masses.
The left one shows the configuration in the complex $\Sigma$ 
plane
and the right one shows the configuration in the real space.}}
\label{slope}
\end{center}
\end{figure}

One should note 
that 1/2 BPS wall configurations and 1/1 SUSY vacua 
are also solutions of the 1/4 BPS equations for the wall webs. 
To illustrate the situation, we 
consider the $\NF=2$ model which has two vacua 
$\langle 1\rangle $ and $\langle 2\rangle $.
Let us first consider a model with real masses 
($M_2=0$) 
$M={\rm diag}(\tilde m_1, \tilde m_2), 
(\tilde m_1<\tilde m_2)$ and 
assume that the configuration depends on only $x^1$ and 
$\Sigma_2=0$.
We can easily 
see that the wall configuration conserves 1/2 
SUSY given by the projection 
$\Gamma _{\rm w}\varepsilon =\varepsilon $ with 
$\Gamma_{\rm w}$ in (\ref{eq:gamma-proj}), 
and interpolates between a vacuum $\langle 2\rangle 
$ at $x^1\rightarrow \infty $ and a vacuum 
$\langle 1\rangle $ at $x^1\rightarrow -\infty $. 
Its tension is expressed as    
\begin{eqnarray}
 T_{\rm wall}
=c\int ^\infty _{-\infty }dx^1\ \partial_1\Sigma _1
=c\big[\Sigma _1\big]^{+\infty}_{-\infty}=
c(\tilde m_2- \tilde m_1)>0.
\end{eqnarray} 
By use of the rotation (\ref{eq:rotation}) and 
the shift symmetry 
$\Sigma \rightarrow \Sigma +\Delta M {\bf 1}_{\NC},\ 
(M \to M-\Delta M {\bf 1}_{\NF}) $, 
we easily obtain a wall configuration for the case of 
a generic complex masses 
$M  = {\rm diag}(m_1+in_1,m_2+in_2)$ with
$(m_2-m_1)^2 + (n_2-n_1)^2 = (\tilde m_2 - \tilde m_1)^2$. 
Therefore, 
this wall configuration is a 1/2 BPS state 
which preserves 1/2 SUSY given by the projection
$e^{-i\frac{\theta}{2}\gamma^5}e^{-\frac{\theta}{2}\gamma^{12}}
\Gamma_{\rm w}
e^{\frac{\theta}{2}\gamma^{12}}e^{i\frac{\theta}{2}\gamma^5}$.
Note that the angle $\theta$ correlates the phase of the 
mass difference $\tan\theta = (m_2 - m_1)/(n_2-n_1)$. 
We also find that this wall 
configuration is mapped into a straight line segment 
connecting two vacuum points $m_1+in_1$ 
and $m_2+in_2$ in the complex field space of $\Sigma $, 
and the wall in the real space of $x^1$, $x^2$
extends along a straight line perpendicular to that line 
segment in the field space as seen in Fig.~\ref{slope}. 
The tension of the wall per unit length is given by 
\begin{eqnarray}
 T_{\rm wall}=c\sqrt{(m_1-m_2)^2+(n_1-n_2)^2}.
\end{eqnarray}

We can directly derive informations about the wall 
configuration from the solution given by the moduli 
matrix
\be
H_0 = \sqrt c (e^{a_1 + ib_1},e^{a_2 + ib_2}).
\label{mm_wall}
\ee
Notice that $\log\Omega \sim \log\Omega_0$ 
outside the core of the wall. 
Then we observe 
\be
\log\Omega 
&\sim& 
\left\{
\begin{array}{cl}
2(m_1x^1+n_1x^2 + a_1) & 
\text{at\ \ } e^{2(m_1x^1+n_1x^2 + a_1)} 
\gg e^{2(m_2x^1+n_2x^2 + a_2)}\\
2(m_2x^1+n_2x^2 + a_2) & 
\text{at\ \ } e^{2(m_1x^1+n_1x^2 + a_1)} 
\ll e^{2(m_2x^1+n_2x^2 + a_2)}
\end{array}
\right.
.\label{logomega}
\ee
Let us call 
\be
(H_0e^{M_1x^1+M_2x^2})^A=
e^{a_A+m_Ax^1+n_Ax^2}
\label{eq:weight}
\ee
the weight 
of the vacuum $\langle A\rangle$. 
The energy density (\ref{energy}) 
becomes negligible far from the core. 
The wall energy density is concentrated around the 
transition line separating two vacuum domains, 
where two terms in $\log\Omega$ are comparable. 
Namely, the position of the domain wall is determined 
by the condition of equal weights of the vacua 
\be
 (m_1-m_2) x^1 + (n_1-n_2) x^2 + a_1 - a_2 = 0.
\label{wall}
\ee
Thus we confirm that a wall is orthogonal to the vector 
$(m_1-m_2,\,n_1-n_2)$. 

As was mentioned at the beginning of this section, the 
total moduli space of the 1/2 BPS single domain wall 
corresponds to ${\bf C}P^1$.
It is parametrized by the homogeneous coordinate $H_0$ 
given in Eq.~(\ref{mm_wall}). 
From Eqs.(\ref{mm_wall}) and (\ref{wall}) we realize that
a wall configuration is given when we specify a point in 
${\bf C}P^1$.
Namely, generic points in ${\bf C}P^1$ give 1/2 BPS 
configurations of domain walls.
However, not 
all the points of ${\bf C}P^1$ give wall configurations.
There are two special points in ${\bf C}P^1$ manifold
which correspond to vacua $\left<1\right>$ and $\left<2\right>$.
In terms of the moduli matrix the vacuum $\left<1\right>$ 
is given by $H_0 = \sqrt c(1,0)$ and the vacuum 
$\left<2\right>$ by $H_0=\sqrt c(0,1)$.
Both of these can be obtained as limits $a_2 \to -\infty$ 
and $a_1\to-\infty$
from the generic moduli matrix (\ref{mm_wall}), respectively.
Physically these can be understood as follows.
In the limit $a_2 \to -\infty$ the weight 
of the vacuum $\left<2\right>$ vanishes compared to 
that of the vacuum $\left<1\right>$. 
Then the domain wall which divides 
the vacua $\left<1\right>$ and $\left<2\right>$ 
goes to positive infinity (the position of the wall 
$a_1-a_2\rightarrow -\infty$) and
the domain $\left<2\right>$ disappears. 
Similarly, the domain $\left<1\right>$ disappears 
in the limit $a_1\to-\infty$.
Thus we conclude that the generic points of 
${\bf C}P^1 \simeq {\bf R}\times S^1$ 
describe domain walls and the remaining two special points 
are vacua. 
The total moduli space ${\bf C}P^1$ is the union of the 
subspaces of a one wall sector 
and zero wall sectors (vacua) 
\be
{\cal M}_{\rm tot} \simeq {\bf C}P^1
= {\cal M}^{\text{1-wall}}_{1/2}
\ \bigcup\ 
{\cal M}^{\text{
vacua}}_{1/1}.
\ee

\subsection{Wall junction}

\begin{figure}[ht]
\begin{center}
\includegraphics[height=4cm]{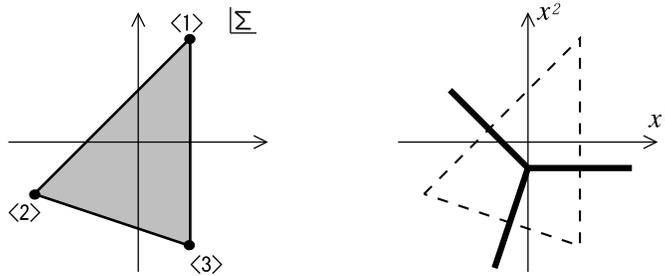}
\caption{\small{The minimal model for 3-pronged wall junction.
The left one is the grid diagram in 
the complex $\Sigma$ plane and the right one is
the web diagram in the configuration space.}}
\label{cp2}
\end{center}
\end{figure}

Let us turn our attention to the case of $\NF = 3$ with 3 
discrete vacua labeled by $\left<A\right>$ $(A=1,2,3)$. 
A 1/4 BPS wall junction firstly appears in this case
since a junction is a soliton which divides 
(at least) 
three domains (vacua). 
Similarly to the 1/2 BPS domain wall it is useful to 
examine the wall junction in the complex $\Sigma$ 
plane. 
While the 1/2 BPS wall is mapped to a line segment 
interpolating two vacua, the 1/4 BPS domain wall junction 
is mapped onto a triangle whose 
three vertices are located at points $m_A+in_A$, 
as shown in the left figure of Fig.~\ref{cp2}. 
We call polygons in the $\Sigma$ plane 
as {\it grid diagrams} 
because very similar diagrams  which are called grid diagrams  
appear in papers \cite{Katz:1997eq,5-brane,Bergman:1998gs} focusing 
the $(p,q)$ string/5-brane webs in the superstring theory.\footnote{
Our polygons in the $\Sigma$ plane do not need to have 
their vertices just on points of the regular 
grid because we are free to choose the mass parameters 
in complex numbers.
However we still use the term, grid diagram, because our 
web has strong similarity
with $(p,q)$ string/5-brane webs.
}
Here we assign the complex masses $m_A+in_A$ so that vacua 
$\langle 1\rangle ,\langle 2\rangle $ and 
$\langle 3\rangle $ are ordered counterclockwise. 
As we have seen in the previous subsection, 
the walls interpolating $\langle A \rangle$ and 
$\langle B \rangle$ extends in $x^1-x^2$ plane along the 
direction perpendicular to the line segment 
$\langle A \rangle \langle B \rangle$ 
of the grid diagram, as shown in Fig.~\ref{cp2}. 

Here we give a comment on boundary conditions in 
solving the master equation (\ref{master}) for 
1/4 BPS states of webs. 
The web has several external legs of walls and then the 1/4 
BPS web becomes asymptotically 1/2 BPS single walls at 
spatial infinities. 
For the case of the 3-pronged junction, 
it is useful to take a limit of the triangular 
boundary whose edges are perpendicular to the external 
legs of the web.
For example, consider a limit of infinite size of the 
dashed triangle in the right 
figure of Fig.~\ref{cp2}.  
On each edge of the boundary triangle we should require 
for the configuration to approach to 
different 1/2 BPS domain walls. 
For more complicated webs which have lots of legs and loops 
as we will deal with in the following, we choose 
an infinite size limit of polygons whose edges are
perpendicular to the external legs as the boundary and require 
the configuration approaches to the 1/2 BPS walls at the edges.

Each component wall $(A,B)=(1,2),(2,3),(3,1)$ 
becomes 1/2 BPS in the spacial infinity 
(along the wall direction) and 
interpolates vacua from $\left<A\right>$ to $\left<B\right>$. 
The wall has the tension  $\vec T^{AB}$ pulling the 
junction along the wall direction outward 
\begin{eqnarray}
 \vec T^{AB}=(Z_2^{AB},-Z_1^{AB}),
\end{eqnarray}
where the central charge of the wall is defined as 
an integral of ${\cal Z}_\alpha$ in Eq.~(\ref{cetral charge}) 
over $-\infty< x^\alpha < \infty$ 
\be
(Z_1^{AB},Z_2^{AB})
\equiv  c(m_B - m_A, n_B - n_A). 
\ee
The magnitude of the tension is 
determined by the mass differences similarly
to the 1/2 BPS wall 
as was shown in the previous subsection. 
We find that these tensions 
balance at the junction to form a static 
configuration, $\sum_{AB}\vec T^{AB}=0$, 
because of the conservation of the central charges 
\be
\sum_{AB} Z_1^{AB} = \sum_{AB} Z_2^{AB} = 0,
\label{pq2}
\ee
where $AB$ runs over the labels of legs which 
extend from the vertex. 
Therefore the junction configuration can be represented 
by the {\it web diagram} (right figure of Fig.~\ref{cp2})
which is obtained by exchanging vertices, edges, and 
faces of the grid diagram with faces, edges, and vertices, 
respectively. 

We can also read the junction charges geometrically 
from the grid diagram. 
Since walls become 1/2 BPS states in the spatial infinity 
(boundaries), the spatial infinity in the $x^1$-$x^2$ 
space are mapped to edges of the grid diagram. 
Therefore the magnitude of the central charge 
$Y$ is given by the area of the triangle
\be
Y = \int dx^1dx^2\ {\cal Y}
= - \frac{2}{g^2}\int_{\triangle} d\Sigma_1 \wedge d\Sigma_2
=-{1\over g^2}(\Delta m_1\Delta n_2-\Delta m_2\Delta n_1)<0.
\ee
where we define 
$\Delta X_i \equiv X_i - X_3, (i=1,2)$ 
for any quantity $X$. 
Notice that the contribution of $Y$-charge of the junction 
to the 
energy is always negative in the case of the Abelian gauge 
theories. 
This negative contribution can be interpreted as a binding 
energy of wall junction.

Let us now examine the moduli of the junction 
\be
H_0 = \sqrt c\left(e^{a_1+ib_1},\ e^{a_2+ib_2},
\ e^{a_2+ib_2}\right).
\label{mm_cp2}
\ee
Positions of 
the junction can be derived by examining 
the asymptotic behavior of $\log \Omega$ as we have done 
for the single domain wall in Eq.~(\ref{logomega}). 
The wall dividing vacua $\left<A\right>$ and 
$\left<B\right>$ sits on a half line 
\be
(m_A-m_B) x^1 + (n_A-n_B)x^2 + (a_A - a_B) = 0, 
\ee
which is consistent with the 
condition of the balance of forces 
given in Eq.~(\ref{pq2}). 
These 
three walls get together at the junction position 
\be
(x^1,x^2) = 
\left(\frac{S_1}{S_3},\frac{S_2}{S_3}\right)=
\left(
\frac{
\Delta n_1\Delta a_2-\Delta n_2\Delta a_1
}{
\Delta m_1\Delta n_2-\Delta m_2\Delta n_1
},
\frac{
\Delta m_2 \Delta a_1 -\Delta m_1\Delta a_2
}{
\Delta m_1\Delta n_2-\Delta m_2\Delta n_1
}
\right),
\label{junction_point}
\ee
where the weights (\ref{eq:weight}) 
of three vacua 
$\langle 1\rangle, \langle 2 \rangle, \langle 3\rangle$ 
become equal each other. 
The vector 
$(S_1,S_2,S_3)=\Delta \vec v_1\times \Delta \vec v_2$ 
is orthogonal to the triangle 
$\{\vec v_1,\vec v_2, \vec v_3\}, (\vec v_A=(m_A,n_A,a_A))$ 
in three dimensions. 
This gives a map from  a real projective space 
${\bf R}P^2$ ($\in {\bf C}P^2$) to the 
$x^1$-$x^2$ space.

For simplicity\footnote{
We can always set $m_3=n_3=0$ by shifting $\Sigma_1$ and $\Sigma_2$
without loss of generality.
}, let us consider the case 
with $[m_A,n_A] = \{[1,0],\ [0,1],\ [0,0]\}$. 
The junction has 
3 external legs of walls: $x^1 = a_3 - a_1$, $x^2 = a_3 - a_2$ 
and $x^2 = x^1 + a_1 - a_2$. 
These 3 walls meet at a point $(x^1,x^2) = (a_3-a_1, a_3-a_2)$.
\begin{figure}[ht]
\begin{center}
\includegraphics[height=4cm]{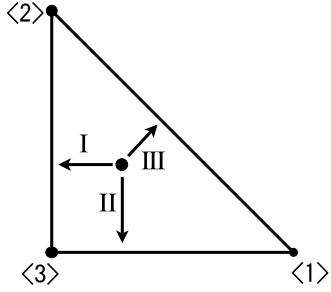}
\caption{\small{The toric diagram for ${\bf C}P^2$ which
corresponds to the total moduli space of the single 
3-pronged junction. 
Points in the face correspond to the 1/4 BPS junctions, 
points in the 
edges to the 1/2 BPS single walls and vertices to the vacua.}}
\label{cp2_toric}
\end{center}
\end{figure}
Generic points in the total moduli space 
${\bf C}P^2$ give 1/4 BPS wall junctions. 

In Fig.~\ref{cp2_toric}, we show the toric diagram\footnote{
By attaching a $U(1)$ fiber to each direction in 
the toric diagram, one obtains the complex manifold, 
${\bf C}P^2$ in this case. 
} 
(triangle) 
of ${\bf C}P^2$, which has three edges corresponding to 
the toric diagrams of three ${\bf C}P^1$'s, respectively. 
The generic points in ${\bf C}P^2$ are brought to the 
boundaries by taking moduli parameters $a_A$ to minus 
infinity.
In this limit the point on ${\bf C}P^2$ moves away from 
the vertex $\left<A\right>$ and finally it get to the edge 
$\left<B\right>\left<C\right>$ ($B\neq C\neq A$).
Physically this is understood as the weight of the vacuum 
$\left<A\right>$ 
vanishes compared to other vacua.
Then the domain $\left<A\right>$ disappears, 
so that the 1/4 BPS 
junction becomes the 1/2 BPS domain wall interpolating 
the vacua $\left<B\right>$
and $\left<C\right>$.

As a concrete example, let us consider the limit 
$a_1\to-\infty$ (we call the limit I). 
In this limit, the junction point $(a_3-a_1,a_3-a_2)$ goes 
away to plus infinity of $x^1$ axis and the only one wall 
which interpolates vacua 
$\left<2\right>$ and $\left<3\right>$ remains at $x^2=a_3-a_2$.
The moduli matrix $H_0$ given in Eq.~(\ref{mm_cp2}) reduces to 
$\sqrt c\left(0, e^{a_2+ib_2}, e^{a_3+ib_3}\right)$. 
This corresponds to the moduli matrix (\ref{mm_wall}) 
for the total moduli space ${\bf C}P^1$ 
of the single 1/2 BPS wall.
\begin{figure}[ht]
\begin{center}
\includegraphics[height=8cm]{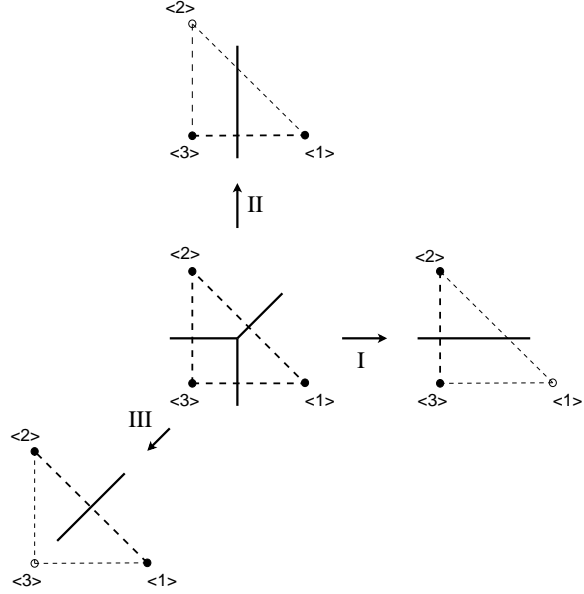}
\caption{\small{Three limits I,II and III of the 1/2 BPS 
3-pronged junction. 
Each limit leads to the different 1/2 BPS walls from the 
1/4 BPS junction.}}
\label{cp2_cp1}
\end{center}
\end{figure}
When we take the other limit $a_2 \to -\infty$ or $a_3 \to -\infty$
(we call the limits II and III) instead of the limit I, 
the other 1/2 BPS walls remain, see Fig.~\ref{cp2_cp1}.
One can take further limit, for example the limit II, 
after taking the limit I, 
where the weight of vacuum $\left<2\right>$ vanishes. 
Then the wall goes away to 
infinity and only the vacuum 
$\left<3\right>$ remains as we showed in the 
previous subsection.

Thus we conclude that 
the total moduli space of the 1/4 BPS 
equation is decomposed 
into the space of the genuine 1/4 BPS junction 
and three subspaces for 1/2 BPS walls 
which conserve different 
half of supersymmetries and three points (SUSY vacua)
as boundaries 
\be
{\cal M}_{\rm tot}
\simeq {\bf C}P^2
= {\cal M}^{\rm junction}_{1/4} \ \bigcup\ 
{\cal M}^{\rm wall}_{1/2} \ \bigcup\ 
{\cal M}^{\rm vacuum}_{1/1}.
\ee

Before closing this subsection, we show an exact solution 
of the wall junction. 
For that purpose, we take 
the gauge coupling squared $g^2$ to infinity. 
The solution is given in Eq.~(\ref{exact}). 
\begin{figure}[ht]
\begin{center}
\begin{tabular}{ccc}
\includegraphics[height=4cm]{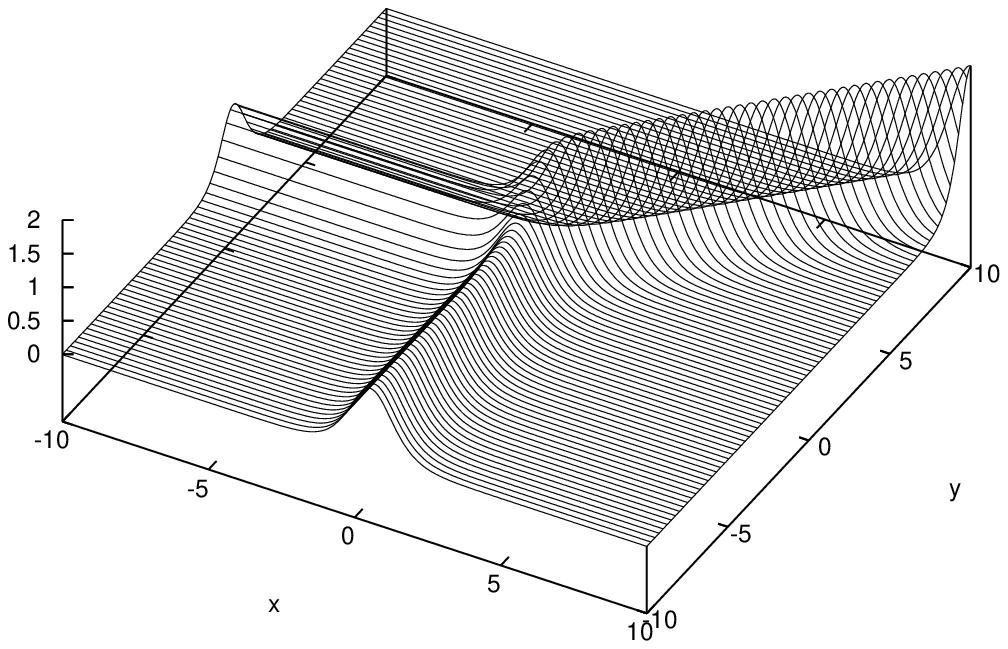} &\qquad\qquad &
\includegraphics[height=4cm]{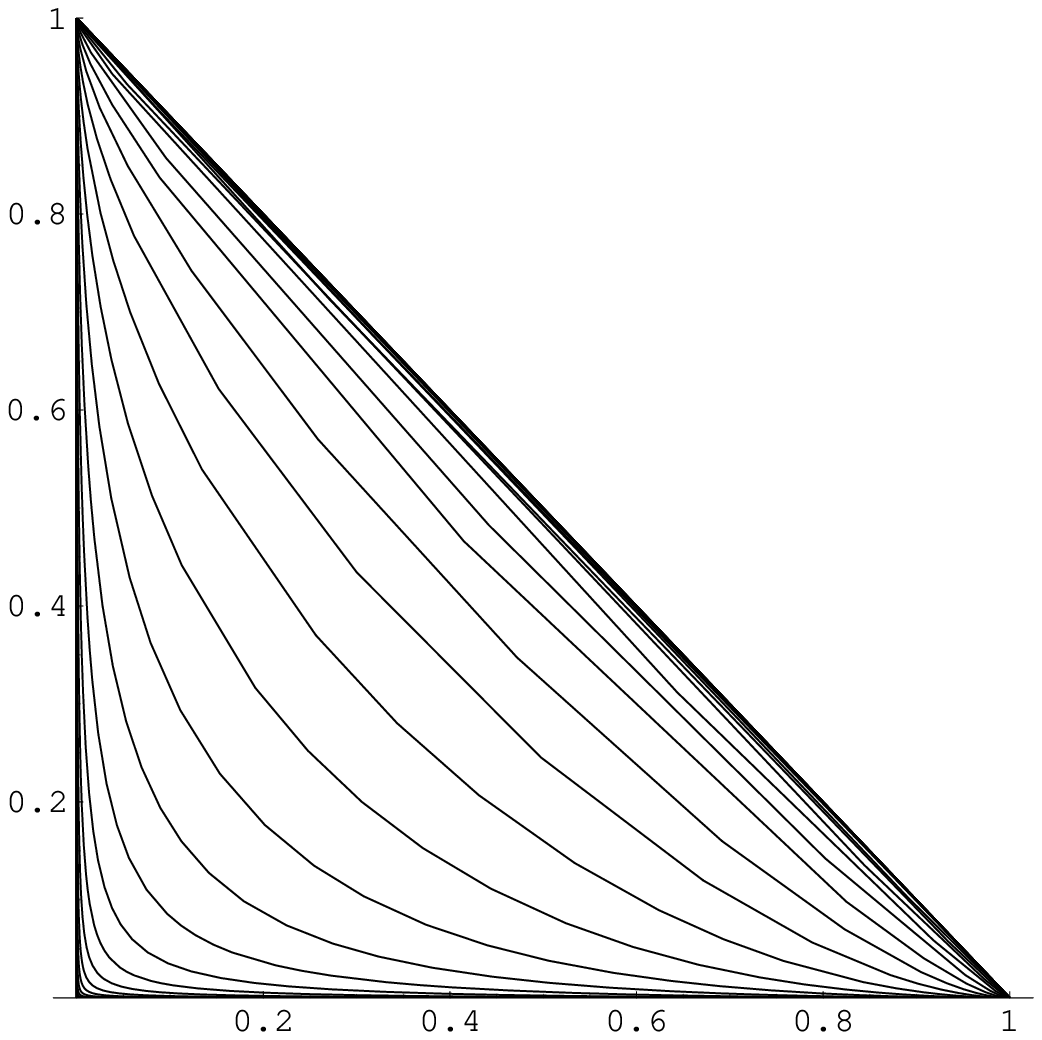} \\
(a) energy density &&(b) the configuration in $\Sigma $ plane 
\end{tabular}
\caption{\small{Exact solution at $g^2\to\infty$.
The model with hypermultiplet masses 
$[m_A, n_A]=[1,0],\ [0,1],\ [0,0]$ 
and with the FI parameter $c=1$. 
We choose the moduli matrix $H_0 = (1,1,1)$. 
In (b), a set of lines $x^1+x^2={\rm const.}$ are mapped to a 
set of curved lines in $\Sigma$ plane.  }}
\label{cp2_exact}
\end{center}
\end{figure}
Fig.~\ref{cp2_exact}~(a) shows the energy density of 
the wall junction in real space and 
Fig.~\ref{cp2_exact}~(b) shows the configuration in 
$\Sigma$ plane. 
The entire 
$x^1$-$x^2$ plane is mapped one-to-one onto the triangle 
in the $\Sigma$ plane (grid diagram).

\subsection{Webs}

Domain wall webs which contain 2 or more wall junctions 
appear in the $U(1)$ gauge theories with $\NF \ge 4$.
The web diagram (in configuration space) 
has $\NF$ faces (domains) corresponding to the 
number of vacua. 
Depending on the values of complex mass parameters 
of the model,
there are two kinds of webs. One is represented by a tree 
diagram and the other is by a diagram with loops. 

\begin{figure}[ht]
\begin{center}
\begin{tabular}{ccc}
\includegraphics[height=4cm]{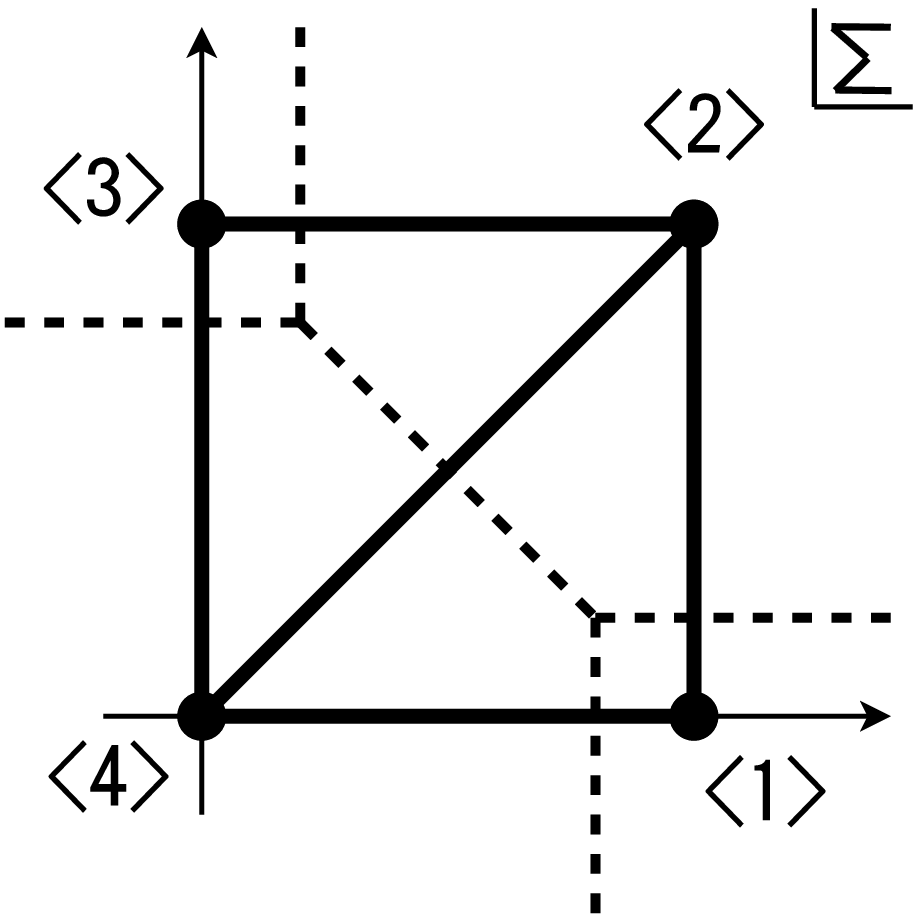} &\qquad\qquad &
\includegraphics[height=4cm]{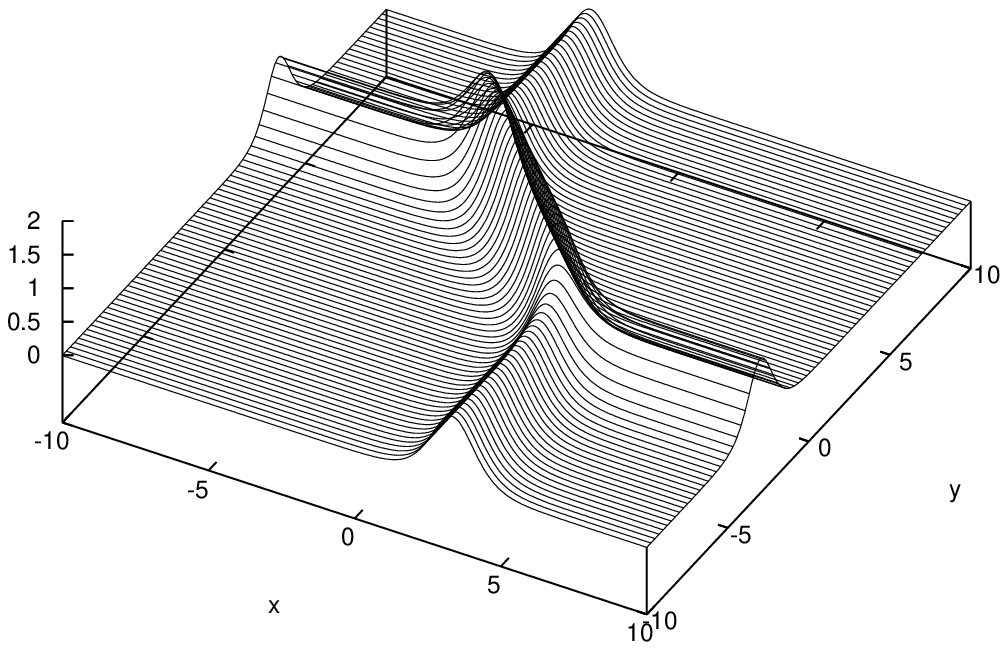} \\
(a) grid diagram &&(b) energy density ($g^2\to\infty$)
\end{tabular}
\caption{\small{The simplest example of the wall web which
has 4 external legs of walls.
The model with hypermultiplet masses 
$[m_A, n_A]=[1,0],\ [1,1],\ [0,1],\ [0,0]$ 
and with the FI parameter $c=1$. 
We choose parameters $(a_1,a_2,a_3,a_4)=(-3,0,-3,0)$. }}
\label{cp3}
\end{center}
\end{figure}

Let us consider the simplest example of wall webs 
in $\NF = 4$ model with
$[m_A,n_A]=\{[1,0],[1,1],[0,1],[0,0]\}$.
The moduli matrix which controls the web is of the form
\be
H_0 = \sqrt c \left( e^{a_1+ib_1},e^{a_2+ib_2},e^{a_3+ib_3},
e^{a_4+ib_4}\right).
\label{mm_cp3}
\ee
These are homogeneous coordinates of the total 
moduli space ${\bf C}P^3$.
The corresponding grid diagram is a
quadrangle in the $\Sigma$ plane 
whose vertices are located at $1,1+i,i$ and $0$. 
Then the web has four external legs. Generally it includes
two 3-pronged wall junctions and an internal leg connecting
these two junctions therein, as in Fig.~\ref{cp3}(a).
The 4 external legs of walls are located at 
$x^1=a_4-a_1$, $x^2=a_4-a_3$,
$x^1 = a_3-a_2$ and $x^2=a_1-a_2$. 

\begin{figure}[ht]
\begin{center}
\includegraphics[height=4cm]{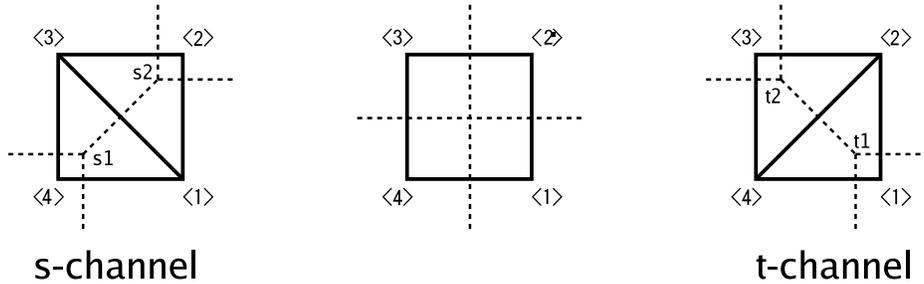}
\caption{\small{The s-channel (left) and t-channel (right) 
in the grid diagram of $N_{\rm F}=4$ case. 
The middle diagram shows the transition point of s- and 
t-channel where two junctions meet.}}
\label{channel}
\end{center}
\end{figure}

This web has two branches which we call s- and t-channel.
The s-channel appears when we choose the moduli parameters 
in the region where $a_1 + a_3 > a_2 + a_4$. 
In this region the web has two junctions, s1 dividing 
vacua 
$
\{\left<4\right>,\left<1\right>,\left<3\right>\}$, 
and s2 dividing vacua 
$
\{\left<1\right>,\left<2\right>,\left<3\right>\}$. 
These are located at $(x^1, x^2)=(a_4-a_1,a_4-a_3)$ 
and $(a_3-a_2,a_1-a_2)$,
respectively. 
The internal wall which connects the junctions 
s1 and s2 
is the wall separating vacua $\left<1\right>$ and 
$\left<3\right>$, as in the left figure of 
Fig.~\ref{channel}. 
The transition from s-channel to t-channel and vise versa 
occurs at the critical point $a_1 + a_3 = a_2 + a_4$.
At that point 
two junctions get together and 
the web has a 4-pronged junction point with 
4 external legs emanating from it. 
The t-channel arises in the region where
$a_1 + a_3 < a_2 + a_4$.
It has two junctions, t1 
dividing vacua
$
\{\left<4\right>,\left<1\right>,\left<2\right>\}$ located 
at $(a_4-a_1,a_1-a_2)$ and 
t2 dividing vacua 
$
\{\left<4\right>,\left<2\right>,\left<3\right>\}$
located at $(a_3-a_2,a_4-a_3)$, respectively.
The internal wall which connects two junctions in 
the t-channel is a domain wall separating vacua 
$\left<2\right>$ and $\left<4\right>$, 
as in the right figure of Fig.~\ref{channel}. 

\begin{figure}[ht]
\begin{center}
\includegraphics[height=4cm]{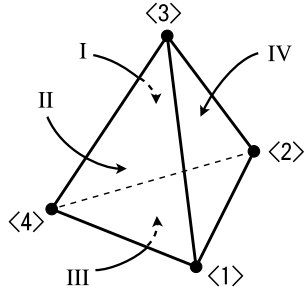}
\caption{\small{The toric diagram for ${\bf C}P^3$ which
is the total moduli space of the web for $\NF=4$ model.
Points in the body, 4 faces and 6 edges correspond to the 
web of 2 junctions, the single junctions and the single 
walls, respectively.
The vertices correspond to the vacua.}}
\label{toric_cp3}
\end{center}
\end{figure}

\begin{figure}[ht]
\begin{center}
\includegraphics[width=11cm]{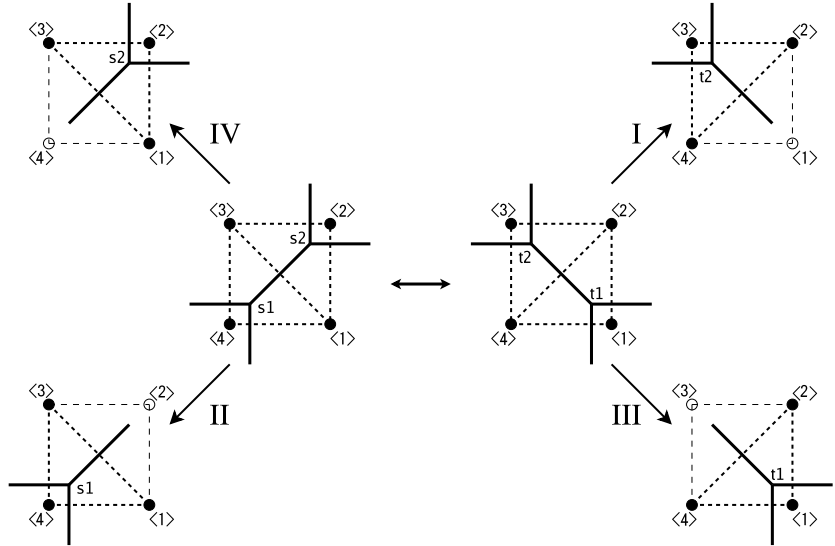}
\caption{\small{There are 4 limits I, II, III and IV where
the web of 2 junctions reduces to the single 3-pronged junction.}}
\label{cp3_cp2_st}
\end{center}
\end{figure}

The total moduli space for this web is ${\bf C}P^3$. 
It is coordinatized by the moduli matrix given 
in Eq.~(\ref{mm_cp3}). 
The toric diagram
for ${\bf C}P^3$ is a tetrahedron. The tetrahedron 
($\langle 1 \rangle \langle 2 \rangle \langle 3 \rangle 
\langle 4 \rangle$) 
has a body, 4 faces ($\langle 1 \rangle \langle 2 \rangle 
\langle 3 \rangle, \cdots$), 6 edges 
($\langle 1 \rangle \langle 2 \rangle, \cdots$), 
and 4 vertices 
($\langle 1 \rangle, \langle 2 \rangle, \langle 3 \rangle, 
\langle 4 \rangle
$), as shown in 
Fig.~\ref{toric_cp3}.
The generic points in ${\bf C}P^3$, namely points in the 
body of the tetrahedron correspond to the webs which have 
two 3-pronged junctions (s-channel or t-channel) in 
Fig.~\ref{channel} as discussed above. 
Let us take the limit of one out of four 
parameters $a_1,a_2,a_3,a_4$ going to $-\infty$, 
which we denote as I,II,III and IV, respectively. 
Then the rank 
of moduli matrix is reduced by one. 
For example, 
let us consider the limit I.
In the limit the junction point t1 at 
$\left(a_4-a_1,a_1-a_2\right)$ is brought to 
the positive infinity along the direction 
$(x^1,x^2) \propto (1,-1)$. 
As a result the configuration has only a single junction t2. 
In fact, the moduli matrix reduces to 
$H_0 = \sqrt c \left( 0,e^{a_2+ib_2},e^{a_3+ib_3},
e^{a_4+ib_4}\right)$
which expresses ${\bf C}P^2$ for a 3-pronged junction, 
as in Eq.~(\ref{mm_cp2}). 
From the viewpoint of the toric diagram of 
${\bf C}P^3$ this limit corresponds to the
procedure where points in the body 
of the tetrahedron 
$\left<1\right>\left<2\right>\left<3\right>\left<4\right>$ 
is taken to points
in the face $\left<2\right>\left<3\right>\left<4\right>$.
By taking other limits II, III and IV, 
we can let points in the body 
$\left<1\right>\left<2\right>\left<3\right>\left<4\right>$
go to points in other faces of ${\bf C}P^3$, as illustrated 
in Fig.~\ref{cp3_cp2_st}. 
After taking one of the limits I, II, III or IV, 
one can further 
take another limit.
For example, let us take the limit II after taking the 
limit I (I$\to$II). 
As a result the junction t2 at $(a_3-a_2,a_4-a_3)$ goes 
to plus infinity of $x^1$ axis. 
Then we obtain the 1/2 BPS wall dividing vacua 
$\left<3\right>$ and $\left<4\right>$, 
as we have shown in the previous subsection. 
The moduli matrix also reduces to that for ${\bf C}P^1$ as
$H_0 = \sqrt c \left( 0,0,e^{a_3+ib_3},e^{a_4+ib_4}\right)$.
Moreover, we can take the limit III or IV after taking the 
limit I and II. 
In the limit I$\to$II$\to$III, for instance, 
only the vacuum state $\left<4\right>$ remains 
and the corresponding 
moduli matrix reduces to\footnote{
Here we used the world-volume 
symmetry (\ref{eq:world-volume-sym}) 
to bring $e^{a_4+ib_4}\rightarrow 1$. 
} $H_0 = \sqrt c \left(0,0,0,1\right)$.

Thus we conclude that 
the total moduli space of the 1/4 BPS wall webs in $\NF=4$ model
is the union of subspaces of a 2-junction sector, 
1-junction sectors, single wall sectors 
and vacua as 
\be
{\cal M}_{\rm tot}
\simeq {\bf C}P^3
= {\cal M}^{\text{2-junctions}}_{1/4} \ \bigcup\ 
{\cal M}^{\text{1-junction}}_{1/4} \ \bigcup\ 
{\cal M}^{\rm wall}_{1/2} \ \bigcup\ 
{\cal M}^{\rm vacuum}_{1/1}.
\ee

\begin{figure}[ht]
\begin{center}
\begin{tabular}{ccc}
\includegraphics[height=4cm]{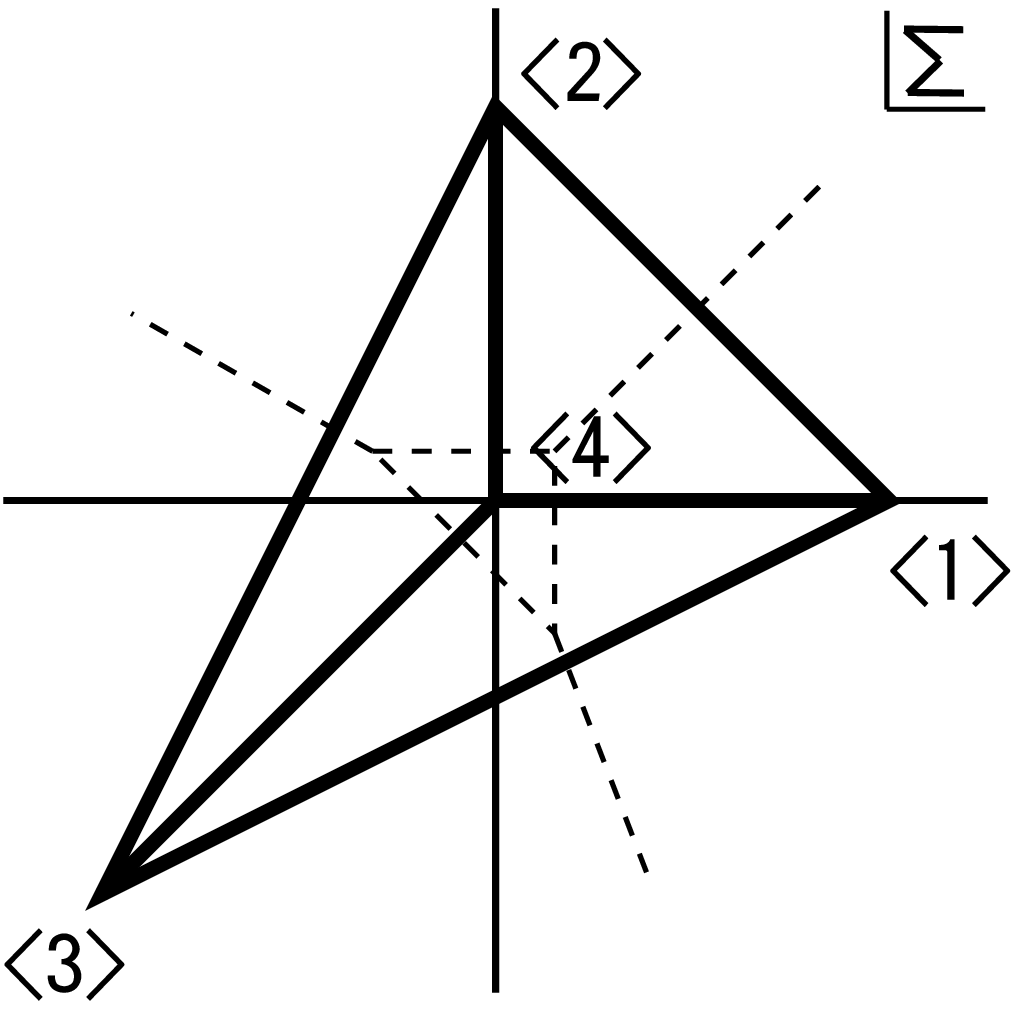} & \qquad\qquad&
\includegraphics[height=4cm]{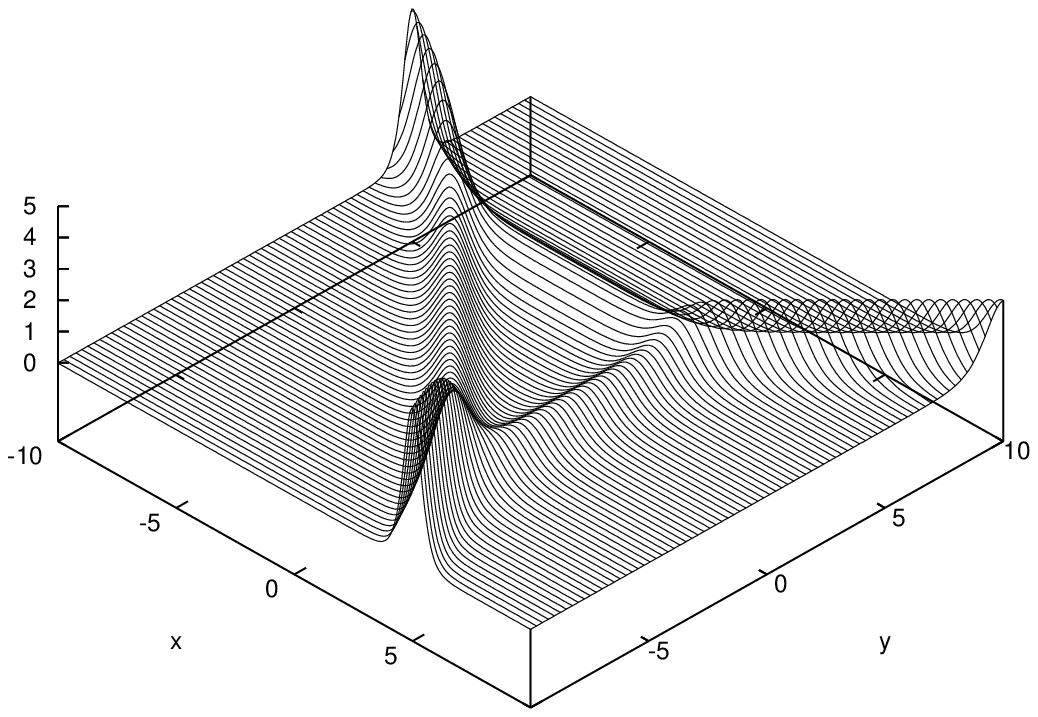}\\
(a) grid diagram &&(b) energy density ($g^2\to\infty$)
\end{tabular}
\caption{\small{Another example of the web in the $\NF=4$ model.
The web has 3 external legs and 1 loop.
The model with hypermultiplet masses 
$[m_A, n_A]=[1,0],\ [0,1],\ [-1,-1],\ [0,0]$ 
and with the FI parameter $c=1$. 
We choose parameters $(a_1,a_2,a_3,a_4)=(-1.5,-1.5,-1.5,1.8)$. }}
\label{cp3_loop}
\end{center}
\end{figure}

\begin{figure}[ht]
\begin{center}
\includegraphics[height=7.5cm]{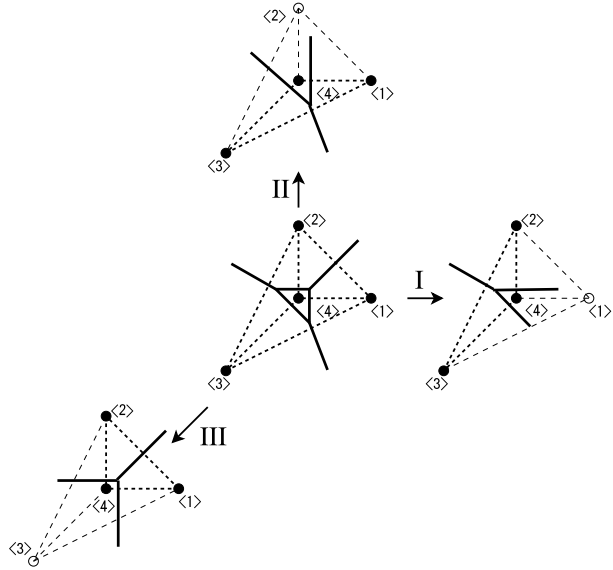}
\caption{\small{The 3 limits I, II, and III,  
where 2 out of 3 junctions in the loop
are brought to infinity. }}
\label{cp3_cp2_loop}
\end{center}
\end{figure}

For different choices of complex mass parameters in 
the $\NF=4$ model, 
we can have another type of the wall web which has a loop. 
In Fig.~\ref{cp3_loop} we show the wall web for the model 
with $[m_A,n_A] = \{[1,0],[0,1],[-1,-1],[0,0]\}$. 
This web is described by the moduli matrix 
in Eq.~(\ref{mm_cp3}). 
The web has three external legs of walls.
Similarly to the s- and t-channel 
of the tree diagram, 
this web has also two branches. 
One branch has a loop with 3 external legs attached 
and another branch has only a single 3-pronged 
junction without loops. 
The loop branch arises in the region $a_1+a_2+a_3 < 3a_4$ and
has three 3-pronged junctions which divides three domains
$\{\left<4\right>,\left<3\right>,\left<1\right>\}$, 
$\{\left<4\right>,\left<1\right>,\left<2\right>\}$ and
$\{\left<4\right>,\left<2\right>,\left<3\right>\}$,
as shown in Fig.~\ref{cp3_loop}.
Their positions are given by
$(a_4-a_1,a_1+a_3-2a_4)$, $(a_4-a_1,a_4-a_2)$ and 
$(a_2+a_3-2a_4,a_4-a_2)$, respectively.
Similarly to the case of the tree type wall web,
we can reduce the loop web to a single 3-pronged junction 
by taking a limit where one out of three vacua $\left<1\right>$,
$\left<2\right>$ and $\left<3\right>$ is taken away to infinity, 
as in Fig.~\ref{cp3_cp2_loop}.
Namely, take one of $a_1$, $a_2$ and $a_3$ to minus infinity 
(we denote the limit I, II and III), respectively.
From the viewpoint of the toric diagram of ${\bf C}P^3$, 
these limits correspond 
to letting points in the body 
of the tetrahedron 
$\left<1\right>\left<2\right>\left<3\right>\left<4\right>$ 
to three faces 
$\left<4\right>\left<2\right>\left<3\right>$,
$\left<4\right>\left<3\right>\left<1\right>$ and 
$\left<4\right>\left<1\right>\left<2\right>$, respectively.

\begin{figure}[ht]
\begin{center}
\includegraphics[height=4cm]{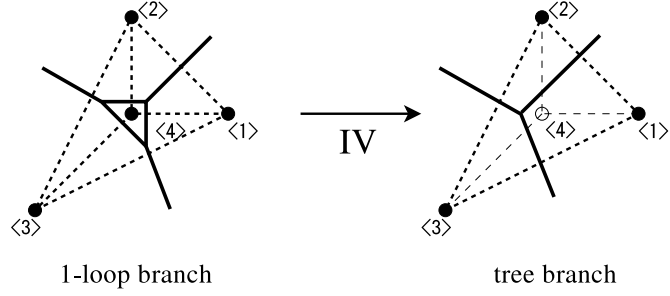}
\caption{\small{By letting 
 $a_1+a_2+a_3 > 3 a_4$, the loop in the web 
shrinks and reduces to the 3-pronged junction. 
In the limit IV with $a_4 \rightarrow -\infty$, 
the effect of vacuum $\langle 4\rangle$ disappears.  }}
\label{loop_tree}
\end{center}
\end{figure}

The loop branch 
can make a transition at the critical point 
$a_1+a_2+a_3 = 3a_4$ to another 
branch with a single 3-pronged junction. 
At that point three 3-pronged junctions in the loop branch get
together and the loop shrinks to a junction 
which divides three domains 
$\{\left<1\right>,\left<2\right>,\left<3\right>\}$. 
In the region $a_1+a_2+a_3 > 3 a_4$, 
the vacuum $\langle 4\rangle$ becomes invisible and 
the parameter $a_4$ has almost no effects on the energy 
density of the web, as illustrated in Fig.~\ref{loop_tree}. 
The position of the junction is determined by only 
three parameters $a_1,a_2,a_3$ as 
$((a_2+a_3-2a_1)/3,(a_1+a_3-2a_2)/3)$.

When we take $a_4$ to minus infinity (we denote 
the limit by IV), effect of the 
vacuum $\left<4\right>$ disappears and the web
reduces to the 3-pronged junction completely. 
Combining this limit IV with the other limits I, II and 
III, we find four faces of the tetrahedron. 
Thus we conclude that the total moduli space
${\bf C}P^3$ of $\NF=4$ loop web is the union of the 
subspace of several topological sectors as
\be
{\cal M}_{\rm tot}
\simeq {\bf C}P^3
= {\cal M}^{\rm loop}_{1/4} \ \bigcup\ 
{\cal M}^{\text{1-junction}}_{1/4} \ \bigcup\ 
{\cal M}^{\rm wall}_{1/2} \ \bigcup\ 
{\cal M}^{\rm vacuum}_{1/1}.
\ee

\begin{figure}[ht]
\begin{center}
\includegraphics[height=3cm]{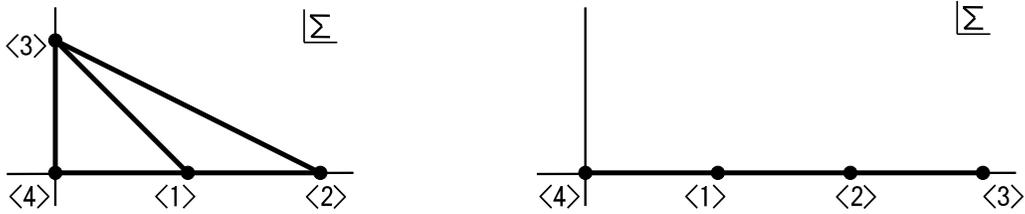}
\caption{\small{Grid diagrams for the special cases where
the mass parameters partially or completely parallel. 
In these case the web has parallel walls therein or 
consists of parallel walls only.}}
\label{cp3_sc_spe}
\end{center}
\end{figure}

There are special cases where the mass parameters are 
partially or completely parallel. 
In such cases a part or whole of the web consists of 
parallel walls. 
We show the web with masses 
$[m_A,n_A] = \{[1,0],[2,0],[0,1],[0,0]\}$ and 
$[m_A,n_A] = \{[1,0],[2,0],[3,0],[0,0]\}$ in Fig.~\ref{cp3_sc_spe}. 

\begin{figure}[ht]
\begin{center}
\begin{tabular}{cc}
\includegraphics[width=6cm]{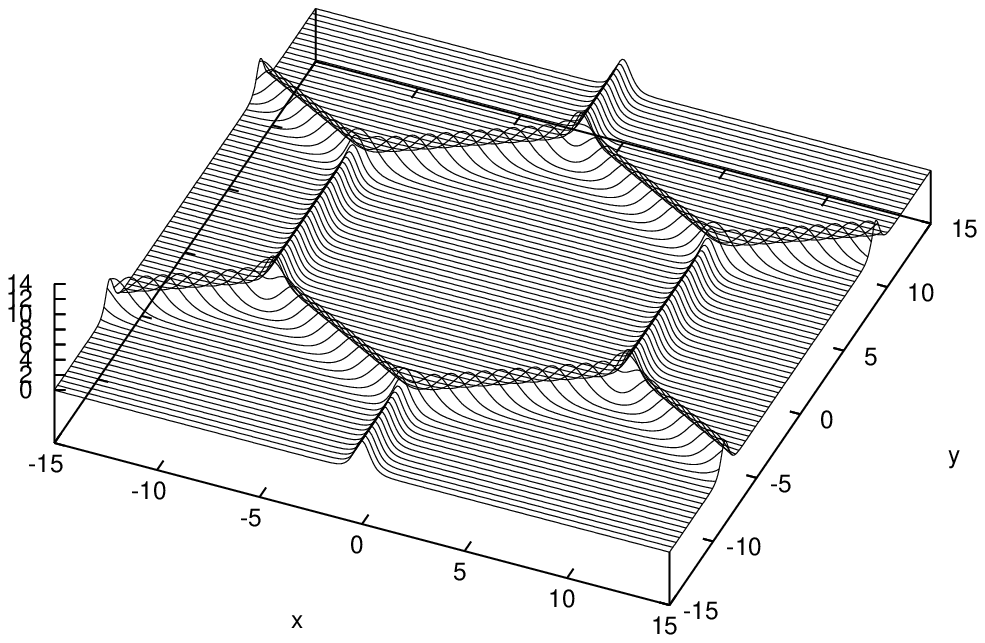} &
\includegraphics[width=6cm]{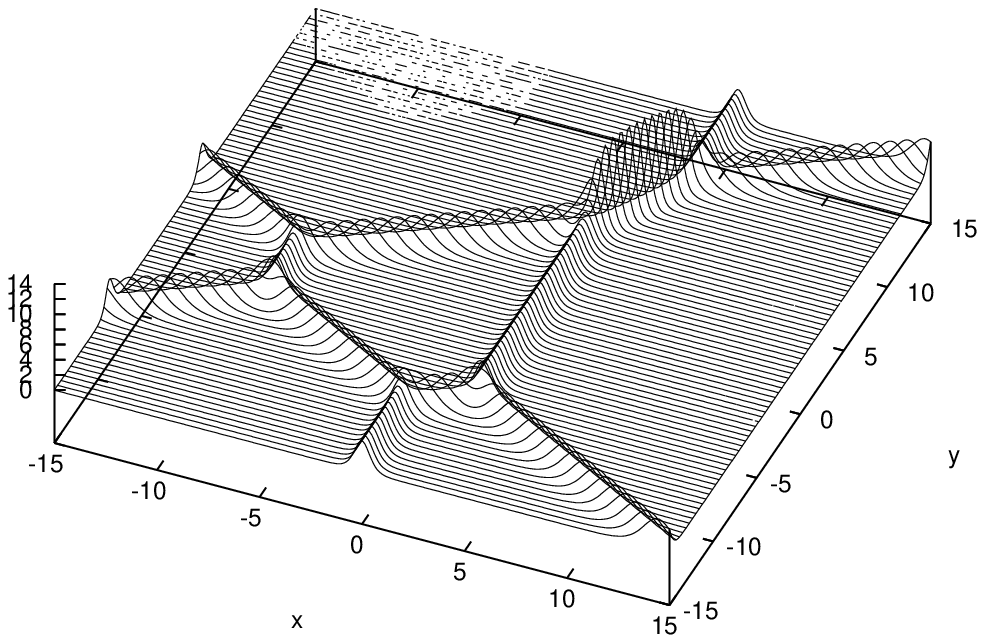}
\end{tabular}
\caption{\small{Mass parameters are chosen as 
$[m_A, n_A]=[2,0],\ 
[1,\sqrt 3],\ 
[-1,\sqrt 3],\ 
[-2,0]
$, $[-1,-\sqrt 3],\ [1,-\sqrt 3]$ and
$[0,0]$.
The left one is the regular hexagon with parameters
$(a_1,a_2,a_3,a_4,a_5,a_6,a_7)=r(1,1,1,1,1,1,0)$ and
the right one has $(r/3,-16,r/3,r,r,r,0)$ with $r=-10\sqrt 3$.}}
\label{webs}
\end{center}
\end{figure}

Next let us consider general cases of $N_{\rm F}$. 
Suppose we have a grid diagram for the $\NF-1$ case.  
If we add one more flavor with a mass outside 
of the grid diagram, the number of edges of the grid diagram, 
that is, the number of external legs in the web diagram 
increases by one. 
If we add a flavor with a mass inside of the grid diagram,
then three internal legs are added 
to the web diagram and those form a loop.
Therefore 
the graphical relation for the web diagram is given by 
\be
\NF = F = E_{\rm ext} + L,
\label{graphical_relation}
\ee
where 
$F$ is the number of faces,
$E_{\rm ext}$ is the number of external legs and 
$L$ is the number of loops in the web, respectively.
Conversely, this relation implies that there are just 
enough degrees of freedom to shift external legs and to shrink 
loops. 
By removing one of the junctions or reduces one of 
the loops to a vertex, we obtain configurations 
corresponding to $N_{\rm F}$ number of 
${\bf C}P^{\NF-2}$'s as boundaries of ${\bf C}P^{\NF-1}$. 
In Fig.~\ref{webs} we show examples of webs which contain 
multiple junctions. The web in Fig.~\ref{webs} divides 7 vacua
and its total moduli space is ${\bf C}P^6$. 
More complicated web is shown in Fig.~\ref{hachinosu} 
with 37 vacua and 
with the total moduli space 
of ${\bf C}P^{36}$. 

\section{Wall Webs in More General Gauge Theories  \label{sect4}}

So far we have considered the 
$U(1)$ gauge group with identical charges for all the 
hypermultiplets as the simplest case. 
As a result we have obtained 
a 3-pronged junction as the fundamental wall junction. 
In this section we discuss wall webs in 
more general gauge theories and show that 
multi-pronged fundamental junction can exist.
To this end, we introduce a method to understand 1/4 BPS 
junctions from the viewpoint of the 1/2 BPS parallel walls. 
For simplicity, we restrict ourselves to the case 
of the $U(1)$ gauge theory 
to explain such a method.  
As was mentioned in Sec.~\ref{sect2}, 
our technique of solving 
the 1/4 BPS equations (\ref{bps_eq1}) and 
(\ref{bps_eq2}) by the moduli matrix $H_0$ is valid for any values
of mass parameters $m_A + in_A$. 
Regardless of mass assignments, the configurations are 
controlled by the same moduli matrix 
\be
H_0 = \sqrt c \left(e^{a_1+ib_1},e^{a_2+ib_2},
\cdots,e^{a_{\NF}+ib_{\NF}}\right),
\ee
and the position of wall which 
interpolates vacua $\left<A\right>$ and $\left<B\right>$ 
can be estimated by comparing the weight of the vacua as 
\be
m_A x^1 + n_A x^2 + a_A \simeq m_B x^1 + n_B x^2 + a_B.
\label{ap_weight}
\ee
This simple formula to determine the approximate 
positions of walls is applicable to both 
the 1/2 BPS parallel walls and the 1/4 BPS wall webs. 
The only difference between them is the slope of the walls 
in the $x^1$-$x^2$ plane.  
Namely, the 1/2 BPS configurations have only 
parallel walls, 
while the 1/4 BPS configurations have walls with different 
slopes which have to meet at some point in $x^1$-$x^2$ plane 
to form a junction.  

We can understand properties of wall webs approximately 
by making use of the similarity between the 1/4 BPS states 
and the 1/2 BPS states as follows. 
When we slice a configuration of 1/4 BPS wall web along 
a line $x^2 =$ constant, for instance, the wall configuration 
on that line is very close (although not identical in detail) 
to the corresponding 1/2 BPS wall configuration 
in the theory with real masses $M_1$ 
[$M_2 \equiv (n_1,\cdots,n_{\NF}) = 0$], 
provided moduli 
are taken to be $x^2$ dependent ($a_A+n_A x^2$ and 
$a_B+n_B x^2$). 
This is because positions of walls in a 1/4 BPS wall web
can be estimated only from the weights of 
vacua as in Eq.~(\ref{ap_weight}), 
which can be interpreted as 1/2 BPS walls~\cite{INOS1,INOS2} 
with $x^2$ dependent moduli $a_A+n_A x^2$ and 
$a_B+n_B x^2$.  
Recall that we can obtain the imaginary part of masses 
$M_2 = (n_1,\cdots,n_{\NF})$ 
from the five dimensional theory 
with real masses $M_1$, 
by the SS reduction along the $U(1)$ symmetry 
generated by $\sum_A n_A T_A$, with $T_A$ the $U(1)$ symmetry 
for the $A$-th flavor.  
Position of walls in the 1/4 BPS wall web can be 
obtained when the modulus, 
which is a complex partner of $\sum_A n_A T_A$, 
is promoted to depend linearly on $x^2$. 
We may call this point of view as the slicing technique.

\begin{figure}[t]
\begin{center}
\includegraphics[height=4cm]{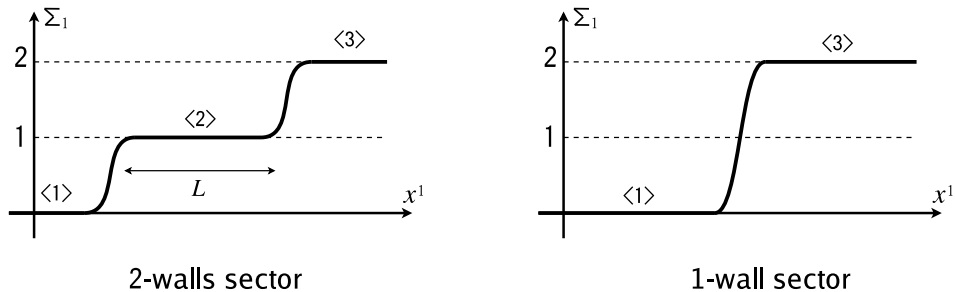}
\caption{\small{The configuration of the adjoint 
scalar $\Sigma_1$ for the 1/2 BPS parallel walls 
is schematically shown as a function of $x^1$.  
The left is two walls sector and the right is one wall 
sector $(a_2 \to -\infty)$.}}
\label{parallel_sigma}
\end{center}
\end{figure}

As a concrete example, let us consider
the real mass assignment ($n_A = 0$) of
the $\NF=3$ model. To avoid inessential complications,  
we set $M = {\rm diag}(0,1,2)$ in the following.
In this case configurations have only parallel walls
which are perpendicular to $x^1$ axis and 
$\Sigma_2=0$ identically.  
Of course, they are 1/2 BPS states.
There are two walls whose positions are estimated as 
$x^1 = a_1 - a_2$ and $x^1 = a_2 - a_3$ for the walls 
connecting $\left<1\right>$ to $\left<2\right>$, 
and $\left<2\right>$ to $\left<3\right>$, respectively. 
In the region where $a_1 - a_2 \ll a_2 - a_3$, namely $a_2$ is 
positive and sufficiently large, we observe two 
walls and the domain of the vacuum $\left<2\right>$ in the middle.  
The relative distance is given by $L = 2a_2 - (a_1 + a_3)$.  
Notice that $L$ has a physical meaning as the relative 
distance\footnote{More precisely 
$L$ should be larger than the width of the wall.
} 
only for $L >0$. 
For $L<0$, the energy density 
becomes almost independent of $L$. 
In the limit $a_2 \to -\infty (L\rightarrow -\infty)$ 
the weight of the vacuum 
$\left<2\right>$ vanishes, and the domain $\left<2\right>$ 
disappears completely. 
The wall configuration $\Sigma_1$ is schematically shown 
for the 1/2 BPS parallel walls in Fig.~\ref{parallel_sigma}.
\begin{figure}[ht]
\begin{center}
\includegraphics[height=5cm]{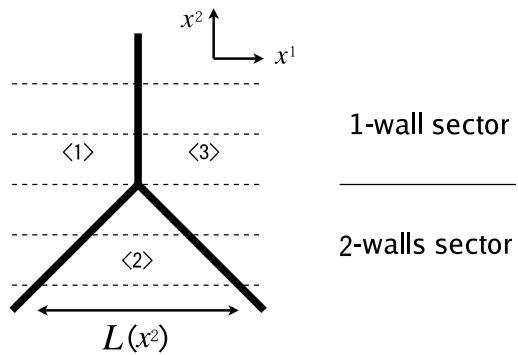}
\caption{\small{The 3 pronged junction of walls. 
The dashed lines indicate 
slices of $x^2 = {\rm const.}$. 
The slices of the configuration can 
be thought of the 1 wall sector above the 
junction point, while the 2 walls sector 
below the junction.}}
\label{slice}
\end{center}
\end{figure}

Let us next turn on the imaginary part of the mass parameters. 
Consider the mass assignment $M = {\rm diag}(0,1-i,2)$.
In this case we get the 1/4 BPS junction which consists 
of three semi-infinite walls. 
Positions of the walls can be estimated from 
Eq.~(\ref{ap_weight}) as $x^1 - x^2 + a_2 - a_1=0$ for the 
wall connecting $\left<1\right>$ to $\left<2\right>$, 
$x^1 + x^2 + a_3 - a_2 = 0$ for $\left<2\right>$ to 
$\left<3\right>$ and $2x^1 + a_3 - a_1 = 0$ for 
$\left<3\right>$ to 
$\left<1\right>$. 
These get together at a junction point 
\begin{equation}
\left({a_1 - a_3 \over 2}, {2a_2 - a_1 - a_3 \over 2}\right). 
\label{eq:junction-point}
\end{equation} 
When we slice the 1/4 BPS junction at various 
values of $x^2 = {\rm const.}$ as illustrated in 
Fig.~\ref{slice}, 
the 1/4 BPS junction can be thought of a 
collection of many 1/2 BPS parallel walls. 
In fact, if we regard $x^2$ as a constant, 
the estimated positions (\ref{ap_weight}) of walls in 
the 1/4 BPS web can be interpreted as that for 
1/2 BPS walls. 
On a slice at $x^2={\rm const.}< a_2-(a_1+a_3)/2$, 
we find a wall dividing $\left<1\right>$ 
and $\left<2\right>$ at $x^1 = a_1 - a_2 + x^2$ and a wall 
dividing $\left<2\right>$ and $\left<3\right>$ at 
$x^1 = a_2 - a_3 - x^2$. 
The relative distance between these two walls 
is given by 
\be
L(x^2) = 2 a_2 - (a_1 + a_3) - 2 x^2. 
\label{relative}
\ee
Notice that the distance $L(x^2)$ vanishes at the 
junction point (\ref{eq:junction-point}) 
and the domain of the vacuum $\left<2\right>$ disappears 
at slices above the 
junction point, $L(x^2) < 0$.  
In this way, a change of the slice position ($x^2$ 
in this case) can be regarded as a change of moduli parameter 
corresponding to the relative distance between parallel 1/2 
BPS walls. 
Therefore we conclude that the 1/4 BPS wall webs
can be constructed by promoting the relative distance moduli 
of the 1/2 BPS parallel walls to a linear function of $x^2$
as in Eq.~(\ref{relative}). 

\medskip
{\sl An example.}
The slicing technique discussed above can be applied 
to find the shape of 1/4 BPS wall webs qualitatively 
when we know the 1/2 BPS walls in 
a theory with real masses.
In Ref.~\cite{Eto:2005wf} we constructed 1/2 BPS wall solution 
in the $U(1)\times U(1)$ gauge theory 
with several hypermultiplets with different 
$U(1)\times U(1)$ charges. 
In particular we considered 
the strong gauge coupling limit where the model reduces 
to a hyper-K\"ahler nonlinear sigma model with the target 
space of the cotangent bundle 
$T^* F_n$ over the Hirzebruch surface $F_n$. 
We now apply our method to this model. 

\begin{figure}[ht]
\begin{center}
\begin{tabular}{ccc}
\includegraphics[height=4cm]{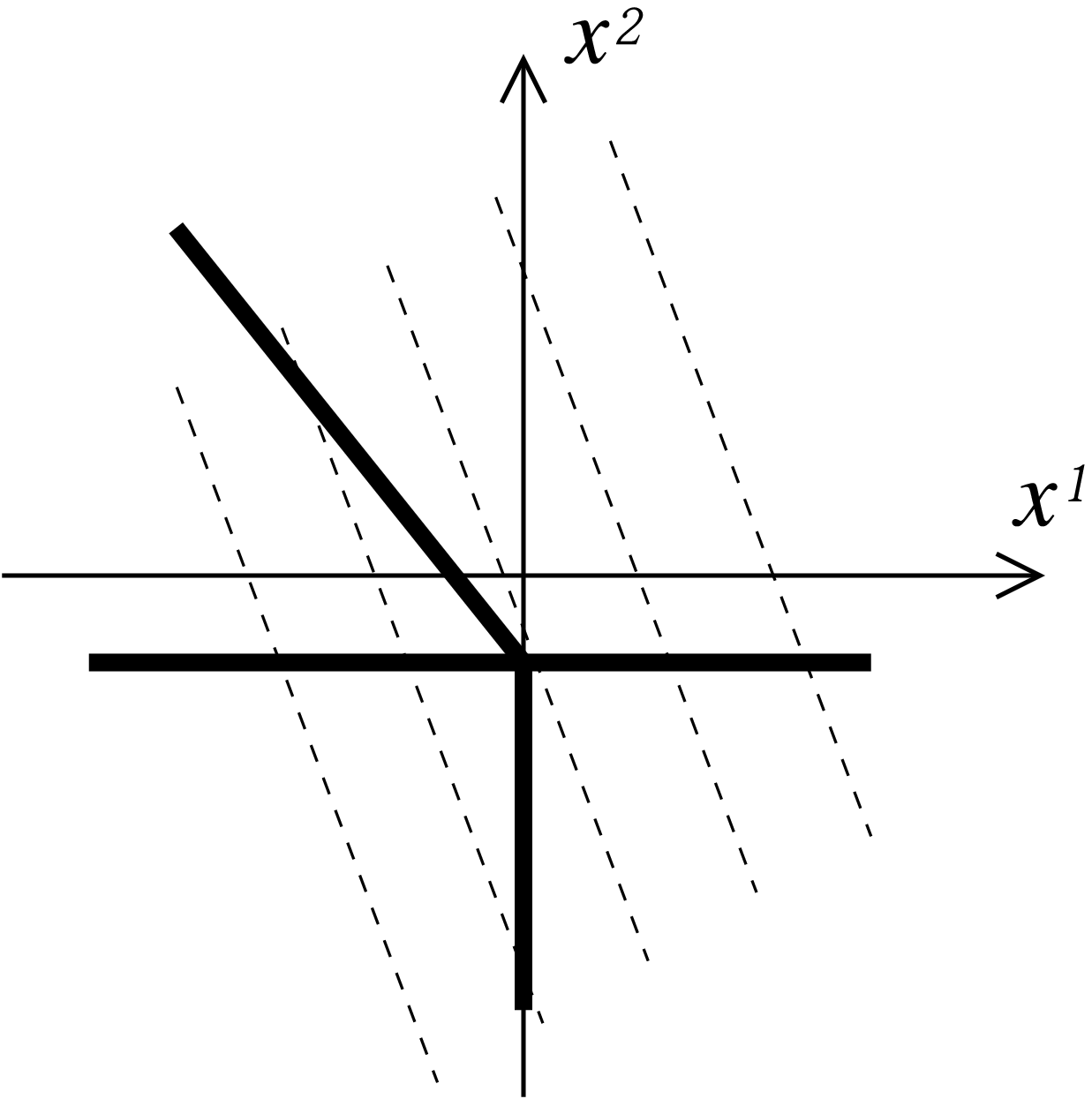} &\qquad\qquad&
\includegraphics[height=4cm]{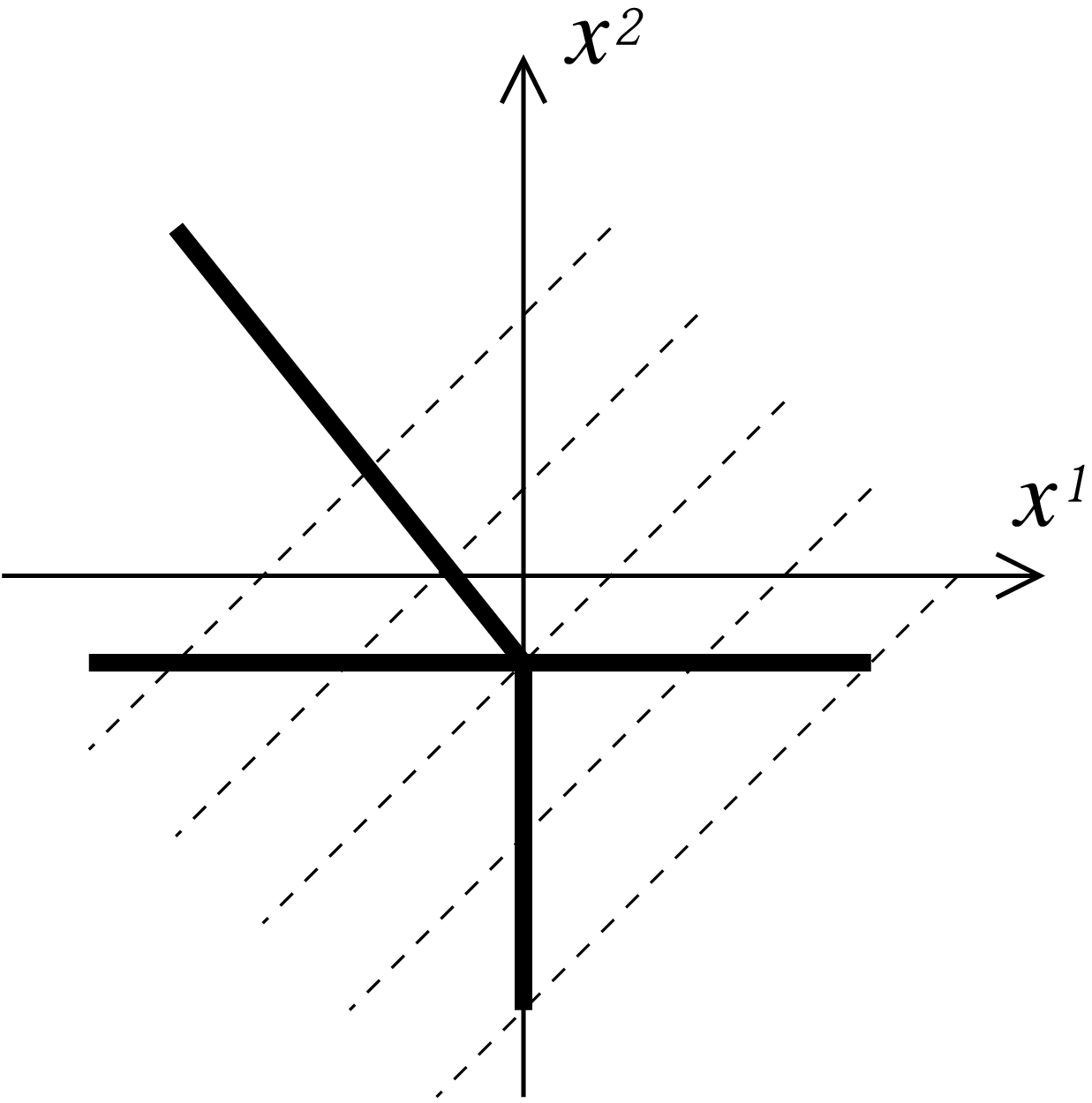} 
\end{tabular}
\caption{\small{Left figure: the slices show the 
three 1/2 BPS parallel walls turning to a single wall. 
Right figure: the} slices show transmutation
of 2 parallel walls.}
\label{fn}
\end{center}
\end{figure}

For one mass arrangement, 
we find that there are only two moduli parameters 
for positions of three parallel 1/2 BPS walls,  
and hence the position of the middle wall is 
locked~\cite{Eto:2005wf}. 
Namely, the relative distance between three walls is 
controlled by only one moduli parameter. 
As the relative distance moduli decreases, three walls 
approach and eventually merge into a single wall. 
According to the slicing technique discussed above, 
we should draw a family of slicing lines (dashes lines in 
the left figure of Fig.~\ref{fn}) progressing to the merging of 3 
walls from the left to a point, and then only single wall 
emerges. 
Thus we find that a fundamental junction is 
not a 3-pronged junction but a 
4-pronged junction in this case, as in the left figure of 
Fig.~\ref{fn}. 

Interestingly we have completely different physics 
for another mass arrangement: 
there exist just two 1/2 BPS walls but 
when they pass through each other 
they transmute to another set of walls with different tensions 
(with total tension unchanged)~\cite{Eto:2005wf}.
In this case also, 
we get the identical 4-pronged junction 
when we promote the relative distance between the two walls 
to a linear function of the coordinate in $x^1, x^2$ 
plane which is perpendicular to the slicing lines, as shown 
in the right figure of Fig.~\ref{fn}. 
Two different physics 
of 1/2 BPS domain walls 
for two different mass arrangements \cite{Eto:2005wf} just 
come from differences of the angles of the slices in the 
same 1/4 BPS wall web.

If a model with real masses 
contains less moduli parameters 
than the number of walls, 
positions of multiple 1/2 BPS walls are 
locked for one mass arrangement. 
When we turn on complex masses in such a model, 
a fundamental wall junction will be 
a multiple-pronged junction 
in the web diagram.

\section{Discussion \label{sect5}}

In this section we discuss possible future directions of work.

{\sl 1. Non-Abelian junction.} 
In this paper we have worked out explicit solutions 
in the Abelian gauge theory and have found 
that the junction charge $Y$ is always negative 
in this case. 
We will show in the subsequent paper~\cite{EINOS2} 
that a positive junction charge $Y$ 
occurs in the non-Abelian gauge theory. 
Only planar diagrams have appeared in 
the $U(1)$ gauge theory as discussed in this paper, 
while non-planar diagrams will appear in non-Abelian 
gauge theory.

\medskip
{\sl 2. Effective field theory on wall webs.}
We can obtain the effective field theory on 
the wall webs using the method of 
Manton~\cite{Manton:1981mp}. 
Let us first recall that the single 1/2 BPS wall has 
a complex moduli parameter associated with the 
broken translation and $U(1)$ flavor symmetries. 
After being promoted to a field on the 
world-volume of walls, the mode function of this complex 
moduli is normalizable when integrated over its single 
co-dimension and 
gives a physical field in the effective theory on the wall.  
However, the webs exist as objects with two co-dimensions and 
their external legs extend as semi-infinite lines.  
Therefore those zero modes which are originally a normalizable 
physical modes of the external legs 
now have semi-infinite support along the external legs. 
Consequently they become non-normalizable modes 
in the effective theory of wall webs.  
In other words, the zero modes which change the boundary 
condition at infinity along the wall direction 
are no longer physical fields in the effective theory of wall 
webs. 
In order to obtain the genuine effective theory on the 
world-volume of the web, 
we should fix such non-normalizable moduli parameters. 
\begin{figure}[ht]
\begin{center}
\includegraphics[height=3cm]{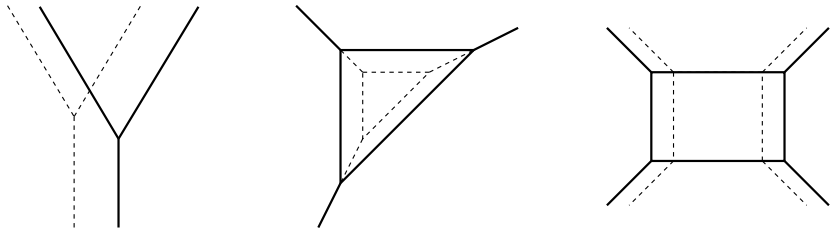}
\caption{\small{Left figure: 
two translational zero (NG) modes. 
Middle figure: a zero mode to change the size of 
a loop. 
This is normalizable and can appear in the effective field 
theory. 
Right figure: a zero mode (QNG) to change the 
shape of the rectangle in one direction. 
This is non-normalizable and appears as a ``coupling 
constant'' in the effective theory.  
}}
\label{zeromodes}
\end{center}
\end{figure}
Zero modes coming from the moduli parameters 
deforming sizes of loops 
are the only possible normalizable modes, 
as seen in the middle figure 
of Fig.~\ref{zeromodes}.
Combined with $U(1)$ zero modes, we thus have $L$ complex 
massless modes if the web contains 
$L = \NF - E_{\rm ext}$ loops with $E_{\rm ext}$ external 
legs, as given in Eq.~(\ref{graphical_relation}).  
We illustrate two examples of the non-normalizable 
zero modes in the left and the right figures of 
Fig.~\ref{zeromodes}. 
The left one of 
Fig.~\ref{zeromodes} describes NG modes 
for broken translational invariance 
along the $x^1$ and $x^2$ coordinates, which 
are accompanied by the NG modes 
for broken $U(1)$ flavor symmetries\footnote{
The non-normalizability of these modes was already 
confirmed in the second reference in 
\cite{Oda:1999az} using the explicit solution.
}.
On the other hand, the right one of 
Fig.~\ref{zeromodes} describes 
the zero mode which changes the shape of the 
rectangular loop. 
Because this mode is not related to any symmetry, 
it is not a NG boson and is called a 
quasi-Nambu-Goldstone (QNG) mode~\cite{QNG}.  
It is important to distinguish the QNG modes from 
the NG modes. 
Since the NG modes for translations have a flat and 
decoupled metric, 
background values of the NG modes 
do not contribute to the effective theory. 
On the other hand, the QNG modes have a nontrivial coupling 
with normalizable modes. 
Namely, the position of the web 
(the left figure of Fig.~\ref{zeromodes}) 
does not enter into the effective theory 
while the shape of the rectangular loop (relative distance between 
left and right sets of external legs) 
appears in the effective theory as ``coupling constants''.

Next we discuss the SUSY in the effective field theory. 
Since the webs are $1/4$ BPS states in 
$d=4$, ${\cal N}=2$ SUSY theory with eight 
supercharges, two supercharges are conserved by 
the $d=2$ effective theory. 
Since the minimum spinor in two dimensions 
is the (one-component) Majonara-Weyl spinor, 
possible SUSY is either 
${\cal N}=(2,0)$ or $(1,1)$. 
Let us recall that the remaining two supercharges 
projected by 
$\Gamma_{\rm w}$ and $\Gamma_{\rm w'}$ in Eqs.~(\ref{eq:gamma-proj}) 
satisfy the $\gamma^0\gamma^3 \varepsilon = \varepsilon$ 
coming from the junction projection $\Gamma_{\rm j}$ 
in Eq.~(\ref{eq:gamma-proj}). 
Since $\gamma^0\gamma^3$ can be regarded as the chiral matrix 
in the $1+1$ dimensions of the $x^0$-$x^3$ plane, 
the remaining two SUSY directions have the same chirality. 
This is also consistent with the fact that 
all of our moduli parameters are complex, because 
the $(2,0)$ SUSY requires scalars in scalar supermultiplets 
to be complex whereas the $(1,1)$ SUSY to be real.  
Since we have obtained the exact solutions in the strong 
gauge coupling limit, 
we may be able to obtain the effective theory explicitly. 
Thus we will obtain a $(2,0)$ nonlinear sigma model~\cite{Hull:1985jv} 
as the effective field theory.
Since this model has a renewed interest 
recently \cite{Witten:2005px},  
it should be worth to pursue 
quantum aspects of the effective theory on the wall webs.

\medskip
{\sl 3. Index theorem.} 
Let us comment on index theorems.  
The master equation (\ref{master}) does not generate 
any additional moduli parameters other 
than those in $H_0$ at least for the $U(1)$ case, 
because of the uniqueness of the solution 
of the master equation shown in Appendix. 
Since the uniqueness of the solution has not been shown 
in non-Abelian gauge theories, 
the index theorem should clarify if there are 
additional moduli. 
Index theorems can count only normalizable modes. 
We can argue in the following that the possible additional 
moduli from the master equation are normalizable 
and are localized around the junction points, if they exist. 
First, 1/4 BPS wall webs always become 
a collection of parallel 1/2 BPS walls 
sufficiently far 
from their junction points. 
Second, 
we know already that all moduli are contained in the 
moduli matrix $H_0$ 
for the 1/2 BPS walls, 
because of the index theorem for 1/2 BPS 
walls~\cite{Sakai:2005sp}.\footnote{
Here we assume non-degenerate masses for hypermultiplets. 
If some masses are degenerate there appear 
non-normalizable zero modes~\cite{INOS2}.
} 
We thus find that the possible additional moduli 
in the master equation (\ref{master}) are normalizable, 
and therefore 
the index theorem for 1/4 BPS wall webs should tell 
whether the moduli matrix contains necessary and sufficient 
number of moduli parameters. 
Namely the index theorem hopefully tells that the wall web 
with $L$ loops has only $L$ complex bosonic zero modes 
as the only additional moduli.  
If we have more zero modes, they should come from solving  
the master equation (\ref{master}). 
This remains as a future problem.

\medskip
{\sl 4. Similarity with $(p,q)$ 5-brane webs.} 
The $(p,q)$ 5-branes in the type IIB string 
theory also can form webs~\cite{5-brane}. 
We have seen that 
our wall webs have many properties in common with 
the $(p,q)$ 5-brane webs. 
Both the wall web and the 5-brane web are 
1/4 BPS states. 
The mass formula for the wall web in terms of complex 
mass differences is very similar to the mass formula for 
5-brane web in terms of RR-charges and NS-NS charges. 
Of course the condition for the balance of force applies 
to both webs.  
Then we reached the grid diagram in the $\Sigma$-plane, 
whose terminology has been borrowed from the $(p,q)$ 5-brane 
webs. 
The dual diagrams of the grid diagrams are webs of 
walls or the $(p,q)$ 5-brane webs, respectively. 
The effective field theory on 5-brane webs is 
$d=5$, ${\cal N}=1$ SUSY gauge theory. 
The light fields in that effective theory 
come from the same zero modes as given in 
Fig.~\ref{zeromodes} ~\cite{5-brane}. 
We thus expect that there should be some similarities  
between $d=2$, ${\cal N}= (2,0)$ sigma models 
and $d=5$, ${\cal N}=1$ SUSY gauge theories. 
This gives a new addition to relations between 
sigma models and gauge theories observed previously.

The phenomenon discussed above 
resembles to a Feynman diagram with 
a 3-point vertex. 
This idea to interpret a web diagram as a Feynman diagram 
has been known for 5-brane webs and is 
called the topological vertex~\cite{Aganagic:2003db}. 
We consider that our model offers an explicit and tractable 
example of the topological vertex in lower dimensions.

\medskip
{\sl 5. Similarity with string webs.}
We see several similarities between wall webs 
and 1/4 BPS dyon as string webs~\cite{1/4dyon,Bergman:1998gs}. 
Another interesting correspondence may be 
the one between monopoles in ${\cal N}=4$ gauge theories 
and walls in ${\cal N}=2$ gauge theories~\cite{Dorey:1998yh}.
The 1/4 BPS dyon can exist in theories with 
gauge group $SU(\NC)$
with $\NC \geq 3$~\cite{1/4dyon,Bergman:1998gs}. 
On the other hand 1/4 BPS wall webs 
can exist if the number of flavors is greater than three, 
$\NF \geq 3$. 
It is known that this kind of flip of gauge/flavor group always 
occurs in relation between monopoles and walls.

\medskip
{\sl 6. Wall lattice.} 
If the number of hypermultiplets is infinite, 
we can obtain the wall webs 
made of infinite number of walls 
which fill the whole space with infinitely many 
junctions. 
Our four dimensional theory can be obtained by 
the Scherk-Schwarz dimensional reduction 
from the theory with massless hypermultiplets 
in six dimensions. 
When we throw away the Kaluza-Klein (KK) modes 
we obtain hypermultiplets with masses 
less than the KK mass.
Whereas if we include all KK towers 
we obtain the infinite number of hypermultiplets 
with the KK 
masses.  
Vacua in this theory becomes points in an 
infinite lattice 
in the $\Sigma$ plane. 
Since a wall can interpolate between a vacuum in 
a fundamental region to a nearest fundamental region,  
we obtain infinitely many wall junctions.
By suitably choosing moduli parameters 
we get a lattice of 
the wall webs with the KK 
masses, which we may call the 
{\it wall lattice}.  
The $SL(2,{\bf Z})$ invariance for torus compactification 
will lead a 
duality on this wall lattice. 
Although these infinite wall webs are interesting in their 
own right, they may become important when one considers 
applications to cosmology.

\section*{Acknowledgements}

This work is supported in part by Grant-in-Aid for Scientific 
Research from the Ministry of Education, Culture, Sports, 
Science and Technology, Japan No.17540237 (N.S.) 
and 16028203 for the priority area ``origin of mass'' (N.S.). 
The works of K.O.~and M.N.~are 
supported by Japan Society for the Promotion 
of Science under the Post-doctoral Research Program  
while the works of M.E.~and Y.I.~are 
supported by Japan Society for the Promotion 
of Science under the Pre-doctoral Research Program. 
M.N.~and N.S.~wish to thank KIAS for their hospitality at the last 
stage of this work. 


\appendix 

\section{Uniqueness of the solution of the master equation} 
In this appendix we show the uniqueness of the master equation 
(\ref{master}) for the case of the Abelian gauge theory
extending the work for walls in Ref.~\cite{SakaiYang}.
In that case $\Omega$ and $\Omega_0$ are positive definite 
functions. 
The master equation can be rewritten as
\be
\left(\partial_1^2+\partial_2^2\right)
 \psi = 1 - \Omega_0 e^{-\psi},\label{master_u1}
\ee
where we redefine the coordinate as 
$(x^1,x^2)\to g\sqrt c(x^1,x^2)$ 
and define $\psi \equiv \log \Omega$. 
Assume that there exist two solutions 
$\psi$ and $\psi'$ of Eq.~(\ref{master_u1}) for a given 
$\Omega_0$ with the identical boundary 
conditions at infinity. 
Then the difference $\delta \psi = \psi - \psi'$ 
satisfies the equation 
\be
\left(\partial_1^2+\partial_2^2\right)
\delta \psi = \Omega_0 e^{-\psi'}
\left(1 - e^{-\delta \psi}\right), 
\label{delta}
\ee
together with the boundary condition 
$\delta \psi\rightarrow 0$ at the spacial infinity.  
Since 
$
(1-e^{-\delta \psi})\delta \psi \ge 0
$, 
we find that 
\begin{equation}
\delta \psi \cdot \left(\partial_1^2+\partial_2^2\right) \delta \psi \ge 0.
\label{eq:positivity}
\end{equation}
Suppose $\delta \psi >0$ at a point $x^1, x^2$. 
Then the boundary condition implies that 
$\delta \psi$ must have a maximum with a positive value. 
At a neighborhood of the maximum, 
$\left(\partial_1^2+\partial_2^2\right) \delta \psi$ 
is always negative, whereas $\delta \psi >0$. 
Thus we obtain 
$\delta\psi \cdot \left(\partial_1^2+\partial_2^2\right) \delta \psi < 0$, 
contradicting (\ref{eq:positivity}). 
Therefore solutions of Eq.~(\ref{delta}) with the boundary 
condition $\delta\psi=0$ 
cannot take positive values anywhere. 
By a similar argument, we find also that 
the solutions cannot take negative values. 
Consequently 
we conclude that $\delta \psi = 0$ is the only solution 
of Eq.~(\ref{delta}) with the boundary 
condition $\delta \psi \rightarrow 0$ at spacial infinity.


\end{document}